\documentclass[a4paper,11pt]{article}
\pdfoutput=1 
\usepackage{jinstpub} 
\usepackage{soul}
\usepackage{lineno,hyperref}
\usepackage{lipsum}
\usepackage{graphicx}
\usepackage{numcompress}
\usepackage{multirow}
\modulolinenumbers[5]
\usepackage{xcolor}
\usepackage{comment}
\usepackage{lineno}
\usepackage{python}

\usepackage{mfirstuc} 
\newcommand{\addReviewer}[2]{
  \expandafter\newcommand\csname #1\endcsname[1]{{\bf \color{#2} \capitalisewords{#1}:\,##1}}
  \expandafter\newcommand\csname #1cor\endcsname[2]{{\color{#2} \capitalisewords{#1}:\,\st{##1}{\bf ##2}}}
  \expandafter\newcommand\csname #1color\endcsname{#2}
}

\addReviewer{alessio}{orange}
\addReviewer{cris}{red}
\addReviewer{evaristo}{blue}
\addReviewer{markus}{BlueViolet}

\def\be{\begin{eqnarray} &&} 
 
\def\ee{\end{eqnarray}}

\title{AI-optimized detector design for the future \\Electron-Ion Collider: the dual-radiator RICH case}

\newcommand{\infnroma}{1}
\newcommand{\sanita}{2}
\newcommand{\lnfn}{3}
\newcommand{\jefflab}{4}
\newcommand{\mitlns}{5}
\newcommand{\howard}{6}
\newcommand{\ferrara}{7}
\newcommand{\cua}{8}
\newcommand{\valparaiso}{9}
\newcommand{\newhampshire}{10}
\newcommand{\northcarolina}{11}
\newcommand{\newmexico}{12}
\newcommand{\oakridge}{13}
\newcommand{\protvino}{14}
\newcommand{\southcarolina}{15}
\newcommand{\illinois}{16}

\newcommand{\olddominion}{17}
\newcommand{\georgiastate}{18}
\newcommand{\losalamos}{19}

\newcommand{\collegenewyork}{20}
\newcommand{\stonybrook}{21}
\newcommand{\williammary}{22}
\newcommand{\abilene}{23}
\newcommand{\argonne}{24}
\newcommand{\duke}{25}

\author[\infnroma,\sanita]{E. Cisbani}
\author[\lnfn]{A. Del Dotto}
\author[\jefflab,\mitlns,*]{C.Fanelli\note[*]{Corresponding author}}
\author[\mitlns]{M. Williams}
\author[\howard]{M. Alfred}
\author[\jefflab]{F. Barbosa}
\author[\ferrara]{L. Barion}
\author[\cua]{V. Berdnikov}
\author[\valparaiso]{W. Brooks}
\author[\newhampshire]{T. Cao}
\author[\ferrara]{M. Contalbrigo}
\author[\northcarolina]{S. Danagoulian}
\author[\newmexico]{A. Datta}
\author[\oakridge]{M. Demarteau}
\author[\protvino]{A. Denisov}
\author[\jefflab]{M. Diefenthaler}
\author[\protvino]{A. Durum}
\author[\newmexico]{D. Fields}
\author[\jefflab]{Y. Furletova}
\author[\southcarolina]{C. Gleason}
\author[\illinois]{M. Grosse-Perdekamp}
\author[\olddominion]{M. Hattawy}
\author[\georgiastate]{X. He}
\author[\losalamos]{H. van Hecke}
\author[\jefflab]{D. Higinbotham}
\author[\cua]{T. Horn}
\author[\olddominion]{C. Hyde}
\author[\southcarolina]{Y. Ilieva}
\author[\cua]{G. Kalicy}
\author[\northcarolina]{A. Kebede}
\author[\collegenewyork]{B. Kim}
\author[\losalamos]{M. Liu}
\author[\jefflab]{J. McKisson}
\author[\valparaiso]{R. Mendez}
\author[\stonybrook]{P. Nadel-Turonski}
\author[\cua]{I. Pegg}
\author[\jefflab]{D. Romanov}
\author[\georgiastate]{M. Sarsour}
\author[\losalamos]{C.L. da Silva}
\author[\williammary]{J. Stevens}
\author[\georgiastate]{X. Sun}
\author[\georgiastate]{S. Syed}
\author[\abilene]{R. Towell}
\author[\argonne]{J. Xie}
\author[\duke]{Z.W. Zhao}
\author[\jefflab]{B. Zihlmann}
\author[\jefflab]{C. Zorn}

\affiliation[\infnroma]{INFN, Sezione di Roma, 00185 Rome, Italy}   
\affiliation[\sanita]{Istituto Superiore di Sanit\`a, 00161 Rome, Italy} 
\affiliation[\lnfn]{Laboratori Nazionali di Frascati, Via Enrico Fermi 40, 00044 Frascati, Italy} 
\affiliation[\jefflab]{Jefferson Lab, Newport News, VA 23606} 
\affiliation[\mitlns]{Laboratory for Nuclear Science, Massachusetts Institute of Technology, Cambridge, MA 02139} 
\affiliation[\howard]{Howard University, Washington, DC 20059} 
\affiliation[\ferrara]{INFN, Sezione di Ferrara, 44100 Ferrara, Italy} 
\affiliation[\cua]{Catholic University of America, Washington, DC 20064} 
\affiliation[\valparaiso]{Universidad T\'ecnica Federico Santa Mar\'ia, Valpara\'iso, Chile} 
\affiliation[\newhampshire]{University of New Hampshire, Durham, NH 03824} 
\affiliation[\northcarolina]{North Carolina A\&T State University, Greensboro, NC 27411} 
\affiliation[\newmexico]{University of New Mexico, Albuquerque, NM 87131} 
\affiliation[\oakridge]{Oak Ridge National Lab, Oak Ridge, TN 37830} 
\affiliation[\protvino]{Institute for High Energy Physics, Protvino, Russia} 
\affiliation[\southcarolina]{University of South Carolina, Columbia, SC 29208} 
\affiliation[\illinois]{University of Illinois, Urbana-Champaign, IL 61801} 
\affiliation[\olddominion]{Old Dominion University, Norfolk, VA 23529} 
\affiliation[\georgiastate]{Georgia State University, Atlanta, GA 30303}  
\affiliation[\losalamos]{Los Alamos National Lab, Los Alamos, NM 87545} 
\affiliation[\collegenewyork]{City College of New York, New York, NY 10031} 
\affiliation[\stonybrook]{Stony Brook University, Stony Brook, NY 11794} 

\affiliation[\williammary]{College of William \& Mary, Williamsburg, VA 2318} 
\affiliation[\abilene]{Abilene Christian University, Abilene, TX 79601}  
\affiliation[\argonne]{Argonne National Lab, Argonne, IL 60439}  
\affiliation[\duke]{Duke University, Durham, NC 27708} 

\emailAdd{cfanelli@mit.edu}

\abstract{
Advanced  detector R\&D requires performing computationally intensive and detailed simulations as part of the detector-design optimization process. 
We propose a general approach to this process based on Bayesian optimization and machine learning that encodes detector requirements.
As a case study, we focus on the design of the dual-radiator Ring Imaging Cherenkov (dRICH) detector under development as a potential component of the particle-identification system at the future Electron-Ion Collider (EIC). 
The EIC is a US-led frontier accelerator project for nuclear physics, which has been proposed to further explore the structure and interactions of nuclear matter at the scale of sea quarks and gluons.
We show that the detector design obtained with our automated and highly parallelized framework outperforms the baseline dRICH design within the assumptions of the current model.
Our approach can be applied to any detector  R\&D, provided that realistic simulations are available.
}


\arxivnumber{1911.05797} 

\begin{document}

\maketitle
\flushbottom

\section{Introduction}\label{sec:introduction}

Applications of methods based on Artificial Intelligence (AI) are becoming ubiquitous in all areas of our society; in particle and nuclear physics in particular, detector Research and Development (R\&D) can certainly benefit from AI and this manuscript provides an explicit example of its usefulness.

In this paper we propose a general approach based on Bayesian optimization (BO)~\cite{jones1998efficient, snoek2012practical} to optimize the design of any detector starting from its R\&D phase.
BO is particularly useful for global optimization of black-box functions that can be noisy and non-differentiable.
The BO approach allows to encode detector requirements like mechanical and geometrical constraints and efficiently maximize a suitable figure of merit used to improve the detector design. 
We apply this general method to a specific use-case: the design of the dual-radiator Ring Imaging Cherenkov (dRICH) detector within the current Electron-Ion Collider (EIC) large scale initiative. 

The future EIC  \cite{accardi2016electron,nas2018} will be a unique experimental facility that will open up new frontiers for exploring Quantum Chromodynamics (QCD), {\em e.g.}\  the 3D structure of nucleons and nuclei in terms of quarks and gluons, and the behaviour of nuclear matter at unprecedented saturated gluon densities. 
A key challenge for EIC detectors involves particle identification (PID)---especially hadron identification---which is required for studying many of the most important experimental processes at the EIC. 
For example, the simulated phase space for pions and kaons in deep inelastic scattering (DIS) processes shows the need for hadronic PID over a momentum range up to 50~GeV/c in the hadron endcap, up to 10~GeV/c in the electron endcap, and up to 5--7~GeV/c in the central barrel~\cite{eichandbook_2019}.

\begin{figure}[t]
\centering
\includegraphics[width=.75\textwidth, angle = 0]{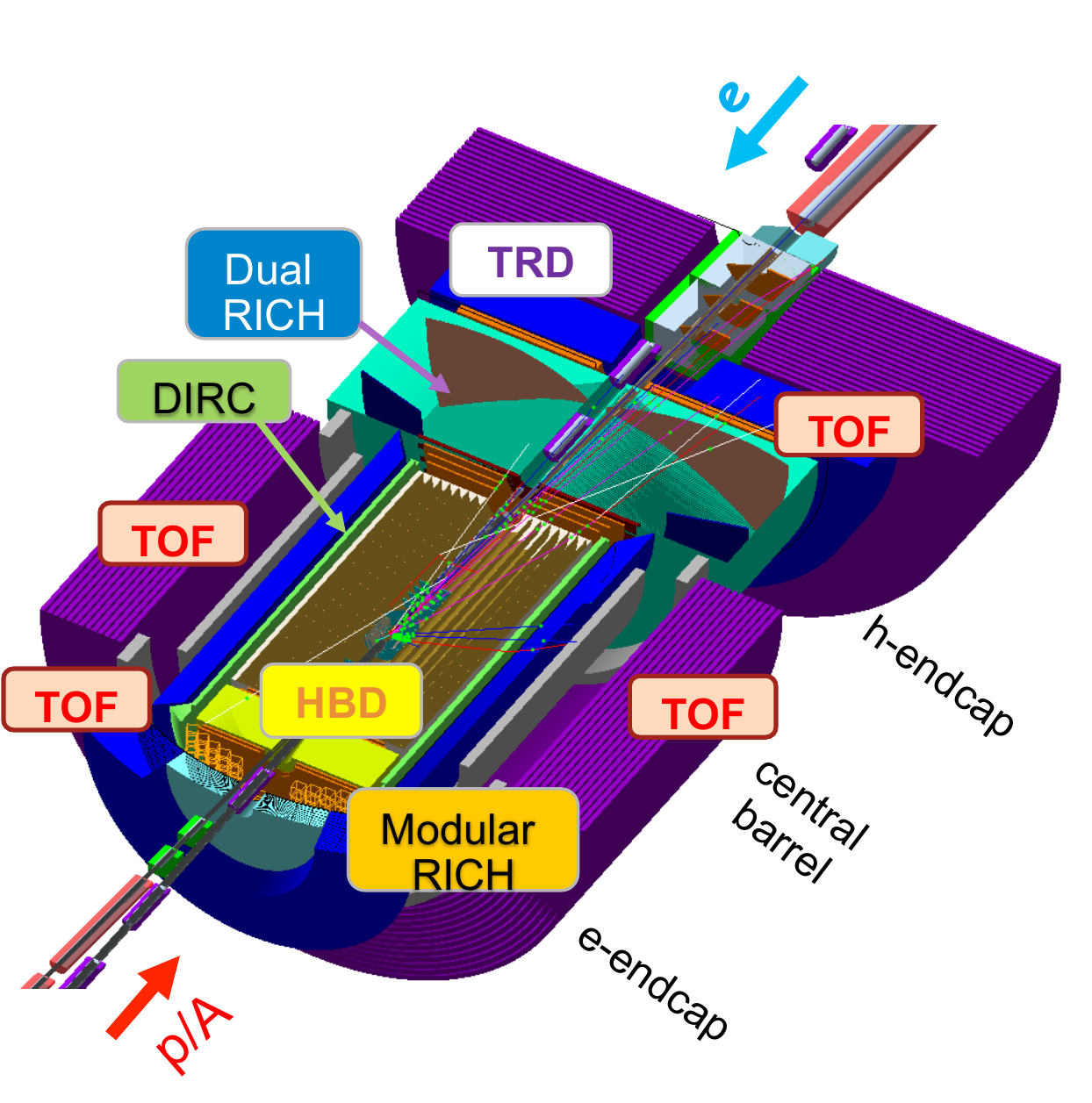}\\
\includegraphics[width=.75\textwidth, angle = 0]{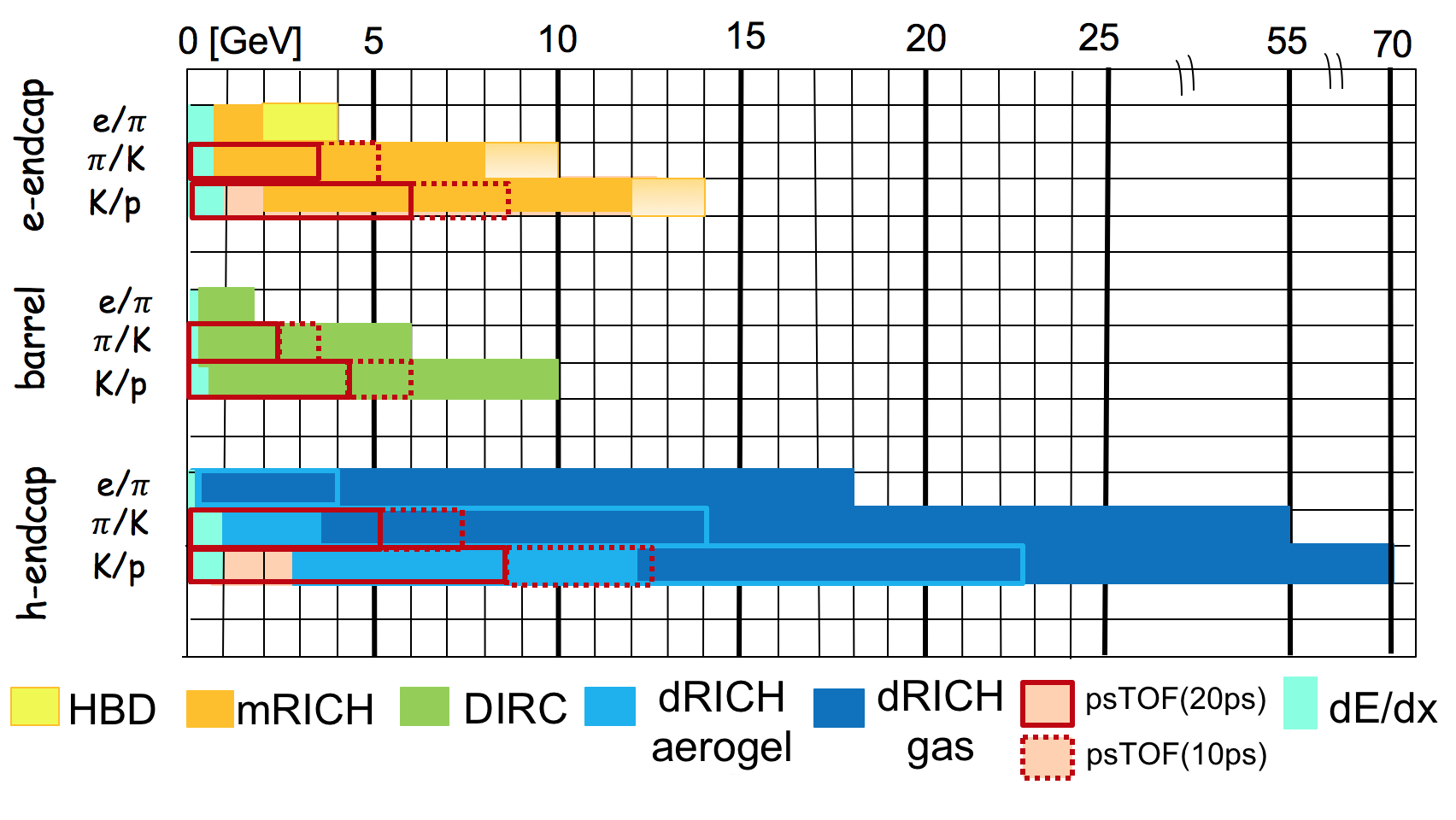}
\caption{(top) 
Detector concept of the Electron-Ion Collider (Jefferson Lab version) showing the location of the dRICH detector and other detection systems.  
(bottom) Projections on the momentum range covered by various PID technologies. 
The spectrometer includes one Hadron Blind Detector (HBD) and the modular RICH (mRICH) in the electron endcap; the dRICH and Transition Radiation Detector (TRD) in the hadron endcap; a large DIRC (Detection of Internally Reflected Cherenkov light) in the barrel and Time of Flight (TOF) detectors everywhere.
Images taken from Ref.~\cite{detector_concept2019}, details about the detectors are summarized in \cite{eichandbook_2019}.
}
\label{fig:mod_perf}
\end{figure}

The purpose of the EIC-PID consortium is to develop an integrated PID system that satisfies
the requirements imposed by the EIC science program.
The current baseline design of the EIC-PID system, shown in Fig.~\ref{fig:mod_perf}, includes the dRICH and a modular-aerogel RICH (mRICH) that uses a Fresnel lens placed in the electron endcap~\cite{del2017design,wong2017modular}. In addition, the Detection of Internally Reflected Cherenkov light (DIRC) detector~\cite{eicdirc} is located in the barrel region, where a fast time-of-flight (TOF) system is foreseen to provide PID for low-momenta particles.

In what follows we show that the new dRICH detector design obtained with our automated and highly parallelized framework based on BO outperforms the baseline dRICH design within the assumptions of the present simulated model.\footnote{Throughout this article, we refer to the dRICH design prior to this work~\cite{del2017design}, as the baseline design.}

\section{Dual-radiator RICH}\label{sec:drich}

\begin{figure}[!ht]
\centering
\includegraphics[scale=0.4, angle = 0]{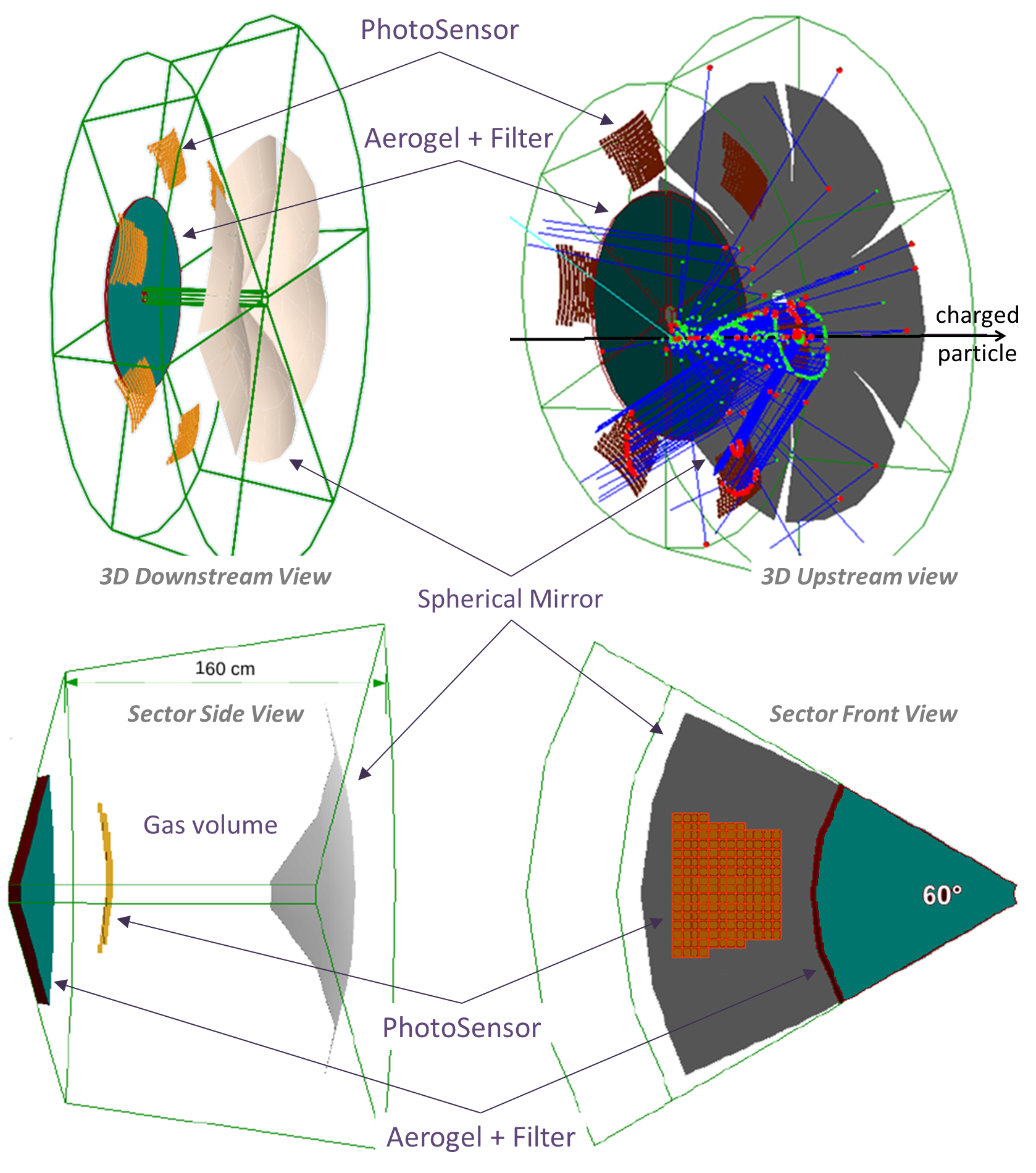}
\caption{ Geant4 model of the dRICH. Top: full 3D downstream (left) and upstream (right) views; bottom: one of the identical 6 sectors, side (left) and front (right) views. The aerogel radiator (disc radius 120 cm, thickness up to 6 cm) and acrylic filter (2 mm thick) are at the detector entrance (colored dark-red and sky blue). The mirrors sectors are close to the exit side, in dark-gray (reflective side) and light-gray (back side). The photon detectors are out of the scattered particle acceptance in orange (sensitive side) and dark-orange (back side) and cover $\sim$4500 cm$^2$ area/sector.  
In the upper/right drawing, a single simulated event (10 GeV/c pion) is represented: the charged particle track is in black while the tracks of the generated optical photons are in blue; the photon reflection and end points in green and red respectively; the large aerogel ring is split into two detectors of adjacent sectors, while the small gas ring is concentrated on one single sensor.}

\label{fig:dual1}
\end{figure}

The goal of the dRICH detector is to provide full hadron identification ($\pi$/K/p better than $3\,\sigma$ apart) from a few GeV/c up to $\approx$50~GeV/c in the (outgoing) ion-side endcap of the EIC detector (see Fig.~\ref{fig:mod_perf}), covering polar angles up to 25~degrees; it will also offer e/$\pi$ separation up to  about 15~GeV/c as a byproduct~\cite{del2017design}.
The dRICH concept was inspired by the HERMES~\cite{akopov2002hermes} and LHCb (RICH1 in Run~1)~\cite{lhcbdetector,adinolfi2013performance} dual-radiator RICH detectors.
The baseline design (see Fig. \ref{fig:dual1}) consists of a large conical-trunk tank ($\sim$160 cm height, $\sim$180$-$220 cm radii) divided into 6 identical, open, sectors (petals); the tank contains an aerogel radiator at the entrance and is filled by C$_2$F$_6$ gas acting as a second radiator. The photons from both radiators share the same outward-focusing spherical mirror and highly segmented ($\approx$3 mm$^2$ pixel size) photosensors located outside of the charged-particle acceptance. 
The baseline configuration is the result of simulation analyses taking into account the geometrical and physical constraints; it can be summarized by the following key parameters: 
i) maximum device length 1.65 m; ii) aerogel radiator refractive index n(400 nm) = 1.02 and thickness 4 cm; iii) C$_2$F$_6$ gas tank length 1.6 m; iv) polar angle coverage $[5^{\circ},25^{\circ}]$; v) mirror radius 2.9 m.

In this study, we benefited from the experience of several groups that have built similar devices in the past~\cite{akopov2002hermes,adinolfi2013performance,alves2008lhcb,nobrega2003lhcb}, and also from the CLAS12 RICH work which is in progress~\cite{pereira2016test}. 
The dRICH baseline detector (see Fig.~\ref{fig:dual1}) has been simulated within a Geant4/GEMC framework~\cite{gemc}.
The Geant4 simulation is based on realistic optical properties of aerogels tested and characterized by the CLAS12 RICH collaboration~\cite{pereira2016test, clas12aerogel}. Absorption length and Rayleigh scattering are included in the simulation, the latter being one of the main sources of background, along with optical dispersion; the spectrum of the Rayleigh scattering is $\propto 1/\lambda^4$, hence this contribution becomes relevant for photon wavelengths below $\sim 300$ nm.\footnote{The $\sim$300 nm value is the safe, approximate, cut-off coming out from the HERMES \cite{akopov2002hermes} and LHCb  \cite{lhcbdetector} studies for their dual-radiator RICHes, where 290 nm and 300 nm have been adopted respectively. The choice of the cut-off is also influenced by the specific dispersion relation of the aerogel; its final value in dRICH will depend from the prototyping outcomes, the aerogel characterization, and likely the application of the optimization method described here.}

\begin{figure}[ht]
\centering
\includegraphics[width=.49\textwidth,angle=0]{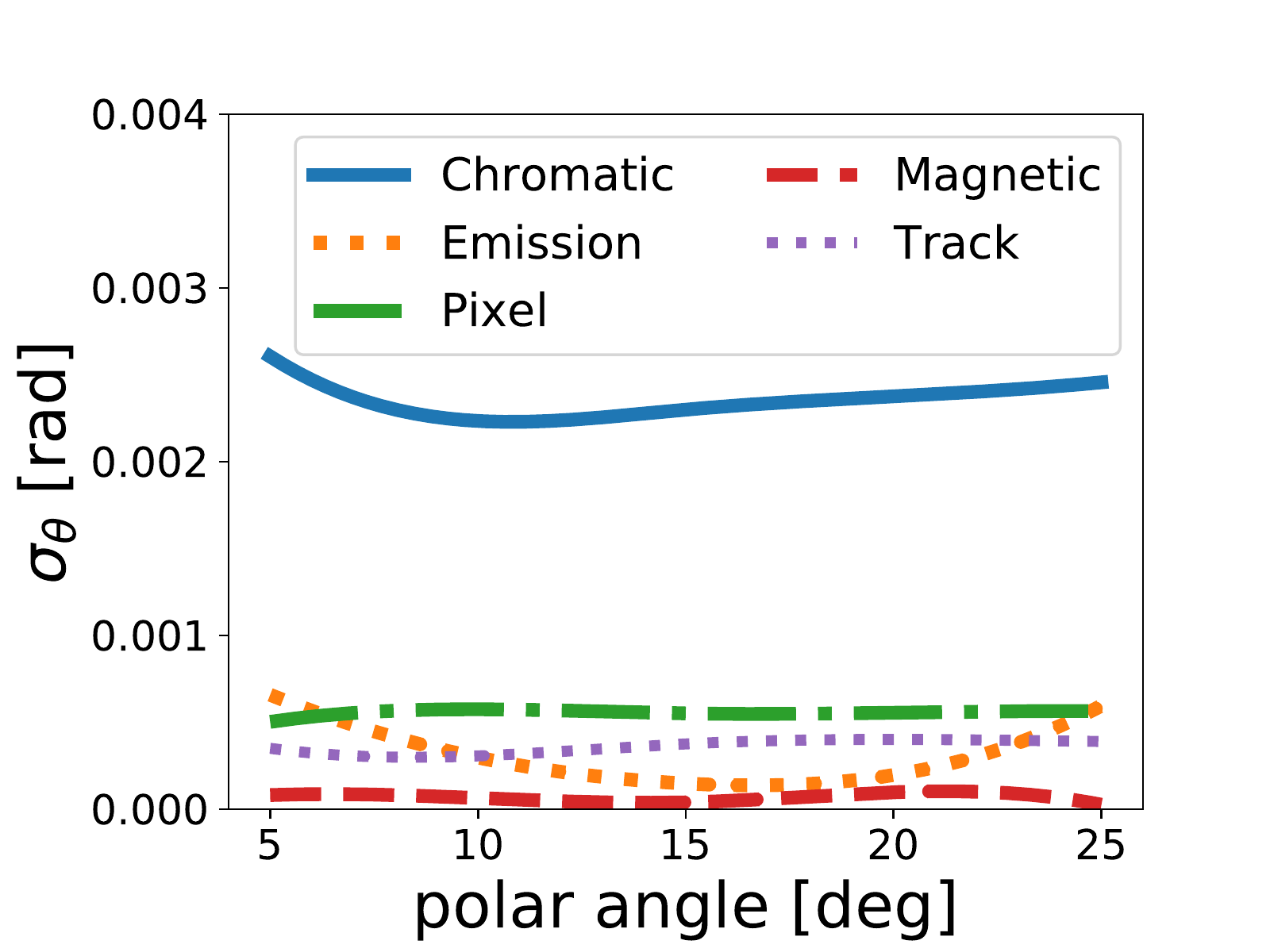}
\includegraphics[width=.49\textwidth,angle=0]{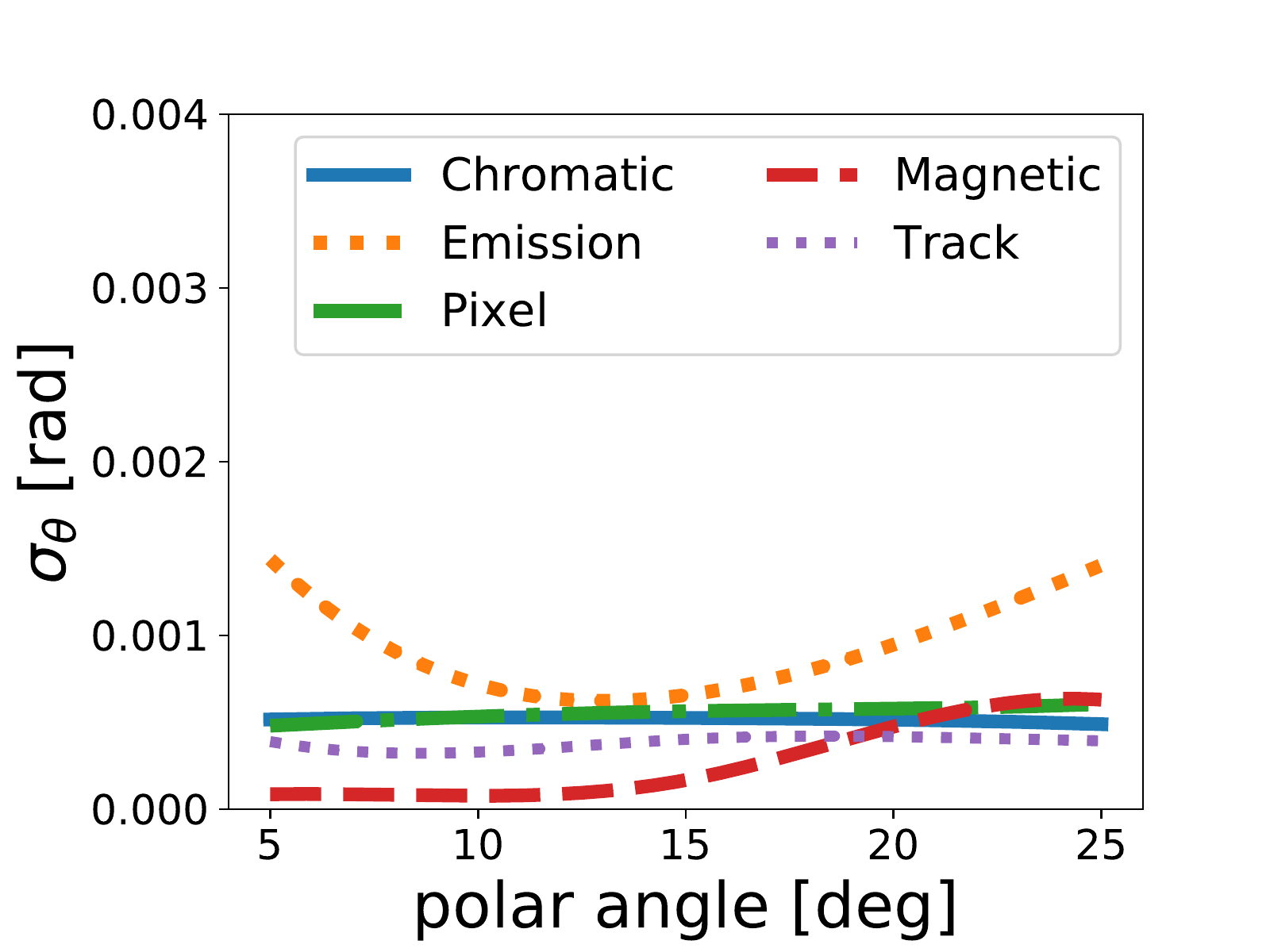}
\caption{
dRICH baseline main contribution to single p.e.\ angular error (see Appendix \ref{app:dispersion} for details on the definition of the error contributions)  for the (left) aerogel and  (right) gas radiators for 30~GeV/c pions, assuming a pixel size of 3~mm. The detector surface has been chosen to minimize the emission uncertainty for central polar angles, namely the detector surface is nearest to the focal surface for central polar angles. 
}
\label{fig:dual2}
\end{figure}

Photons produced in the aerogel with wavelengths below 300~nm are removed, imitating the effects of an acrylic filter that will separate the aerogel from the gas, for both filtering and avoiding chemical degradation of the aerogel. The mirror reflectivity is assumed to be 95\% and uniform.  
The dRICH is in a non-negligible magnetic field and the charged-particle tracks are bending as they pass through the Cherenkov radiators, providing an additional source of uncertainty in the Cherenkov ring reconstruction. The size of this effect is proportional to the path length within the Cherenkov radiators, and therefore, it is important for gas radiator. The magnetic field used in the simulation is that of the JLab detector design at 3T central field.\footnote{The solenoid is still under development and its central magnetic field is now expected to be smaller than 3~T. This value is, therefore, conservative for the dRICH simulation results.  The impact of the magnetic field is confined to the gas angular resolution at large polar angle (right plot of Fig.~\ref{fig:dual2}). Our proposed optimization method and its applicability do not depend on the strength of the magnetic field.} 
The optical sensor area in such a mirror-focusing design can be rather compact, and can be placed in the shadow of a barrel calorimeter, outside of the radiator acceptance. 
The pixel size of the photosensors has been assumed to be 3~mm; the quantum efficiency curve of the multi-anode PMT Hamamatsu-H12700-03 \cite{h12700} has also been assumed, which is similar to that of any potential photosensor suitable for sustaining both relatively high magnetic field and irradiation expected in the dRICH detection region.

The reconstruction of the Cherenkov angles from either aerogel or gas photons coming from the single particle simulated per event is based on the Inverse Ray Tracing algorithm used by the HERMES experiment (see Ref.~\cite{akopov2002hermes} for details on the algorithm).\footnote{The large separation of the Cherenkov photon emission angles in aerogel ($\sim$12 deg) and gas ($\sim$2 deg) makes in practice straightforward the association of the photons hits to the correct radiator.}
Figure~\ref{fig:dual2} shows the main error contributions to the single photoelectron (p.e.)\ angular resolution of the two radiators as a function of the charged particle polar angle, see also appendix \ref{app:dispersion} and Ref.~\cite{del2017design}.

In addition to the angular error, the ring reconstruction is affected by the level of noise, whose main sources are: 
i) the aforementioned aerogel scattered photons which are largely suppressed by a high-pass wavelength filter assumed at 300 nm; this is included in the simulation; 
ii) the intrinsic noise of the sensors and electronics which is simulated as random uniform noise (30 hit/event) in the whole detector area;\footnote{The dRICH sensor and electronics are not defined yet; one can estimate their impact assuming typical performance of current state of the art systems. With 1 kHz sensor dark count rate, in a conventional $\sim$100 ns gated triggered electronics (recent, higher performance, solutions such as streaming readout are expected to be adopted in the final design, the  EIC-eRD23 \cite{eicannual_report_2019} is devoted to this specific topic) we expect a modest $\sim$5 random noisy hit/event/sector, that corresponds to about 3 noisy hits in the aerogel ring area.} 
iii) the physics background related to the event particle multiplicity which is related to the wide EIC physics phenomenology, and is not included in the current simulation.\footnote{The current dRICH development stage is focused on the physical-optical properties of the detector which mainly depend on the expected average phase space of the hadrons.}
In 2019 the dRICH R\&D entered the prototyping phase that, in the coming years, will validate the above described Monte Carlo model and eventually fine-tune the implemented simulator.

\section{Methodology and Analysis}\label{sec:methodology}

Determining how much space to allocate for various detector components, what kind of sensors to use, and which configuration provides the best performance, are often dilemmas that need to be resolved without increasing costs. 
In addition, for multipurpose experiments like the EIC, optimizing the detector design requires performing large-scale simulations of many important physics processes. 
Therefore, learning algorithms ({\em e.g.}, AI-based) can potentially produce better designs while using less resources. 
Nowadays there are many AI-based global optimization procedures ({\em e.g.}\ reinforcement learning, evolutionary algorithm). Among these learning algorithms, the above mentioned BOs have become popular as they are able to perform global optimization of black-box functions, which can be noisy and non-differentiable.
These features make BOs particularly well suited for optimizing the EIC detector design, and, in general, BOs could potentially be deployed for a variety of critical R\&D efforts in the near future. 

BOs search for the global optimum over a bounded domain $\chi$ of black-box functions, formally expressed as $x*=arg_{min \ x  \in \chi} \ f(x)$.\footnote{Here, we assume one is searching for a global minimum. }
 The aim of a BO is to keep the number of iterations required to identify the optimal value relatively small. 
The BO approach has been applied to solve a wide range of problems in different areas from robotics to deep learning~\cite{brochu2010tutorial,eric2008active,brochu2010bayesian,lizotte2007automatic,srinivas2012information,hutter2011sequential,bergstra2013hyperopt}. 
For example, in experimental particle physics it has been used to tune the parameters of Monte Carlo generators~\cite{ilten2017event}. 
When applied to detector design, each point explored corresponds to a different detector configuration (\textit{design point}). 
The function $f$ can be thought of as a figure of merit that has to be optimized ({\em e.g.}, a proxy of the PID performance). Typically Gaussian processes~\cite{williams2006gaussian} (GP) are used to build a surrogate model of $f$, but other regression methods, such as decision trees, can also be used. For example, Gradient Boosted Regression Trees (GBRT) is a flexible non-parametric statistical learning technique used to model very-expensive-to-evaluate functions. The model is improved by sequentially evaluating the expensive function at the next candidate detector configurations, thereby finding the minimum with as few evaluations as possible.
A cheap utility function is considered, called the acquisition function, that guides the process of deciding the next points to evaluate. 
The utility function should balance the trade-off of exploiting the regions near the current optimal solution, and exploring regions where the probabilistic model is highly uncertain.

We have explored the use of BOs in the optimization of the design of the dRICH. 
Each point in the parameter space corresponds to a different detector design,
and consists of a vector of values for the parameters in the Table~\ref{tab:parameters}.
Running the Geant4 simulation
and subsequent analysis of a single detector design point takes about 10~minutes on a single physical core.\footnote{Model name: Intel(R) Xeon(R) CPU E5-2697 v4 @ 2.30GHz, two threads are run per core.} A grid search with just 8 dimensions even run on 50 cores in parallel looks unfeasible due to the ``curse of dimensionality'' \cite{bellman1957dynamic}. 
Several open-source tools are available for Bayesian optimization (see, {\em e.g.}, Ref.~\cite{snoek2012practical}); the results shown in the following are based on the scikit-learn package~\cite{skopt}. 

In this work, eight parameters are considered when optimizing the dRICH design, inspired by the previous studies done in Ref.~\cite{del2017design}: the refractive index and thickness of the aerogel radiator; the focusing mirror radius, its longitudinal (which determines the effective thickness of the gas along the beam direction) and radial positions (corresponding to the axis going in the radial direction in each of the six mirror sectors, see Fig.~\ref{fig:dual1}); and the 3D shifts of the photon sensor tiles with respect to the mirror center on a spherical surface, which contribute to determine the sensor area and orientation relative to the mirror.\footnote{\label{foot:coordinate} The Cartesian reference frame of the tile shifts has the origin in the center of a dRICH conical-trunk sector (see Fig. \ref{fig:dual1}), $z$ is along the beam, $x$ along the sector radii, and $y$ derived accordingly; the single sector detector is then replicated six times to build the whole detector (namely, a standard Geant4 procedure).}
These parameters, reported in Table~\ref{tab:parameters}, cover rather exhaustively the two major components of the dRICH: its radiators and optics. 
They have been chosen 
to improve the dRICH PID performance, 
under the constraint that it is possible to implement any values resulting from the optimization with (at worst) only minor hardware issues to solve. We assume 100 $\mu m$ as the minimum feasible tolerance on each spatial alignment parameter, whereas for the aerogel, we assume 1 $mm$ on the thickness and 0.2\% on the refractive index.  
\\A relevant parameter has been essentially neglected in the optimization: the gas refractive index, whose tuning would require a pressurized detector making this choice hardly convenient.
We also postpone the optimization of the TOF-aerogel transition region, since at the moment these are two separate detectors with different simulation frameworks. 
The parameter space can be extended once detailed results from prototyping and tests will be available.

\begin{table}[t]
\caption{\label{tab:parameters}
The main parameters injected in the Bayesian Optimizer. 
The regions of parameter space explored are based on previous studies~\cite{del2017design}; definition of the parameter reference systems can be found in text and footnote \ref{foot:coordinate}. The tolerances refer to the expected feasible construction tolerances; variation of the parameters below these values are irrelevant.}  
\centering
\scalebox{0.8}{
\begin{tabular}{ | c | c | c | c |}
\hline
\textbf{parameter} & \textbf{description} & \textbf{range} [units] & \textbf{tolerance} [units]\\
\hline
 R & mirror radius & [290,300] [cm] & 100 [$\mu$m]\\ 
 pos r & radial position of mirror center & [125,140] [cm]& 100 [$\mu$m]\\
 pos l & longitudinal position of mirror center & [-305,-295] [cm]& 100 [$\mu$m]\\   
 tiles x & shift along x of tiles center & [-5,5] [cm]& 100 [$\mu$m]\\ 
 tiles y & shift along y of tiles center & [-5,5] [cm]& 100 [$\mu$m]\\
 tiles z & shift along z of tiles center & [-105,-95] [cm]& 100 [$\mu$m]\\
 n$_{\textup{aerogel}}$ & aerogel refractive index & [1.015,1.030]& 0.2\%\\
 t$_{\textup{aerogel}}$ & aerogel thickness & [3.0,6.0] [cm] & 1 [mm]\\
\hline
\end{tabular}
}
\end{table}


Since the aim of the design optimization is to maximize the PID performance of the dRICH, it is natural to build the objective function using the separation power between pions and kaons, defined as
 \begin{equation}\label{eq_nsigma}
    N\sigma = \frac{||\langle \theta_{K} \rangle - \langle \theta_{\pi} \rangle ||\sqrt{N_{\gamma}}}{\sigma^{1 p.e.}_{\theta}},
\end{equation}
where $\langle \theta_K \rangle$ and $\langle \theta_\pi \rangle$ are the mean Cherenkov angles for kaons and pions, respectively, obtained from the angular distributions reconstructed by Inverse Ray Tracing~\cite{akopov2002hermes}. Here, $N_\gamma = (N_\gamma^\pi + N_\gamma^K)/2$ is the mean number of detected photo-electrons  and $\sigma_\theta ^{1p.e.}$ the single photo-electron angular resolution; 
the reconstructed angles are approximately Gaussian distributed, hence $\sigma_\theta^{1p.e.}/\sqrt{N_\gamma}$ corresponds to the averaged RMS of the above mean angles, i.e. the ring angular resolution.

In order to simultaneously optimize the combined PID performance of both the aerogel and gas parts in the dRICH, two working points have been chosen based on the performance of the baseline design. 
In particular, we chose one momentum value with $\approx 3\sigma$ $\pi/K$ separation each for the aerogel and the gas: $p_{1}=$ 14 GeV/c and $p_{2}=$ 60 GeV/c, each close to the end of the curves reported in Fig.~\ref{fig:dual2} for the aerogel and gas, respectively.
The goal here is to optimize the quantities ($N\sigma$)$_{1,2}$ corresponding to the working points $p_{1}$ and $p_{2}$. 
One choice of the figure of merit which proved to be effective is the harmonic mean 
\begin{equation}\label{eq:fom}
\begin{split}
    &h = 2 \cdot \left[ \frac{1}{(N\sigma)|_{1}}+\frac{1}{(N\sigma)|_{2}} \right]^{-1}.\\
\end{split}
\end{equation}
The harmonic mean tends to favor the case where the PID performance ({\em i.e.}\ $N\sigma$) is high in both parts of the detector.
The optimization consists of finding the maximum of the figure of merit (FoM), as defined in \eqref{eq:fom}.\footnote{BOs in skopt search for a minimum, hence we change the sign of \eqref{eq:fom}.}
Samples of pions and kaons have been produced with Geant4~\cite{gemc} using particle guns with central values of the momentum corresponding to 14 and 60 GeV/c. 
\\In order to cover a larger region of the phase-space in the forward sector of the detector, the polar angles have been spanned uniformly from 5 to 15 degrees.

\begin{figure}[t]
\centering
\includegraphics[scale=0.367, angle = 0]{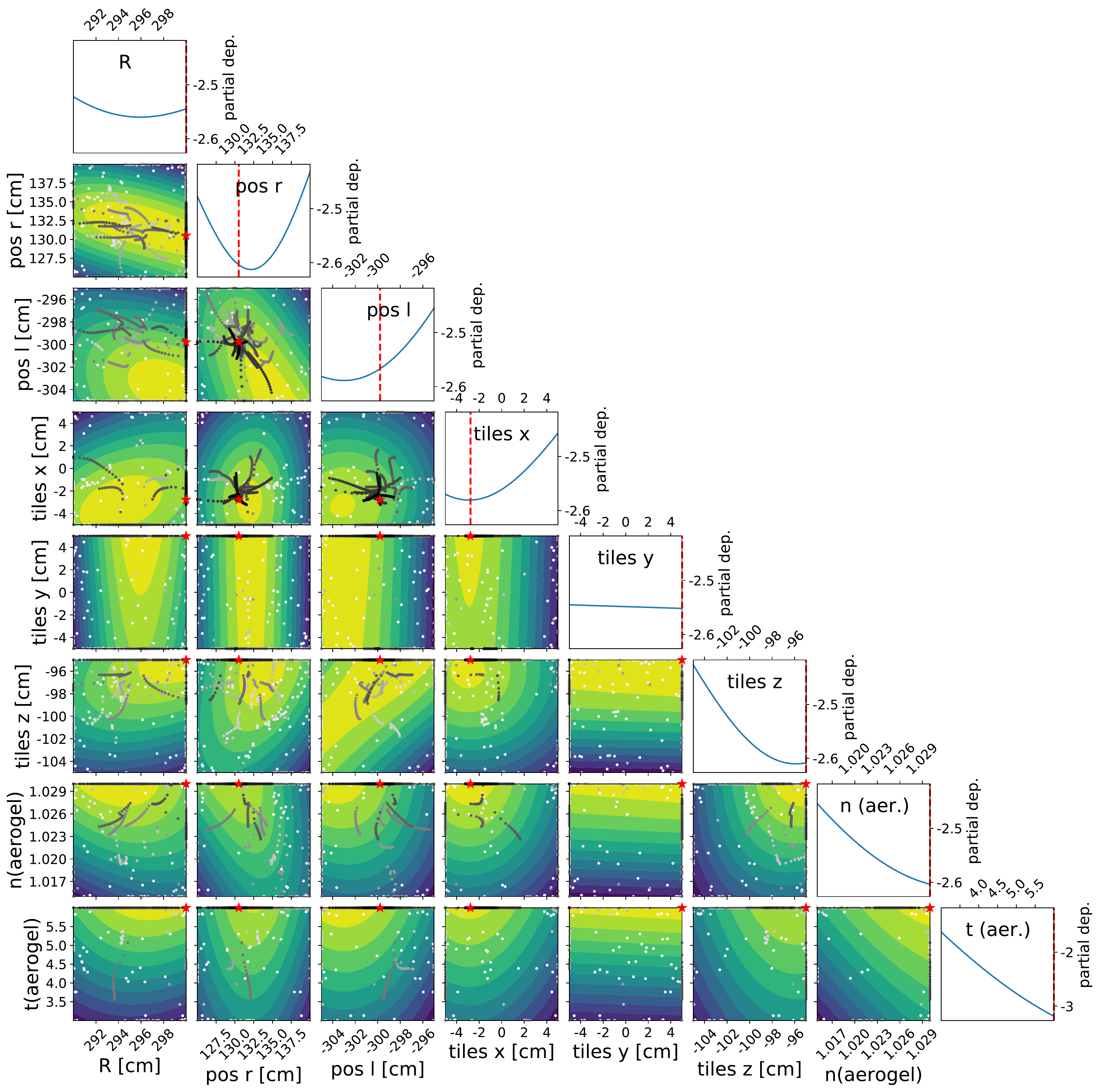}
\caption{2D plot of the objective function (color axis) defined as in Eq.~\eqref{eq:fom}. The optimization strategy of the dRICH design involves tuning the 8 parameters described in Table~\ref{tab:parameters}.  
In order to study possible correlations, each parameter is drawn against the other. The evaluations made by the optimizer are shown through an intensity gradient in the point trail ranging from white (first call of parallel observations) to black (last call). After about 55 calls, the stopping criteria are activated. At the top of each column the partial dependence of the objective function on each variable is shown separately, defined by Eq.~\eqref{eq:partialdep}. Intuitively, the partial dependence provides the importance ranking of each  parameter. The dotted red lines correspond to the projections on each variable of the optimal point found by the BO.
}
\label{fig:posterior}
\end{figure}

Figure~\ref{fig:posterior} shows the posterior distribution after $T$ calls projected in 2-dimensional subspaces of the design parameter space. These plots illustrate the possible correlations among the parameters. The optimal point in each subspace is marked with a red star. Notice that the black points, corresponding to the points evaluated by the BO in its ask-and-tell procedure, tend to form basins of attraction around the minimum. 
Recall that the black-box function we are optimizing is noisy in the Geant4 simulation (see also Appendix~\ref{sec:noise_studies} for a detailed study on statistical uncertainties). 
Plots of the partial dependence of the objective function on each parameter are displayed in Fig.~\ref{fig:posterior}. Given a function $f$ of $k$ variables, the partial dependence of $f$ on the $i$-th variable is defined as~\cite{friedman2001greedy}
\begin{equation}\label{eq:partialdep}
\phi(\theta_{i})=\frac{1}{N}\sum_{j=0}^{N-1}f(\theta_{1,j},\theta_{2,j},\cdots,\theta_{i},\cdots,\theta_{k,j}).
\end{equation}
The above sum runs over a set of $N$ random points from the search space, and the partial dependence 
can be considered as a method for interpreting the influence of the input feature $\theta_{i}$ on $f$ after averaging out the influence of the other variables. 

\begin{figure}[!]
\centering 
\includegraphics[width=.45\textwidth,origin=c,angle=0]{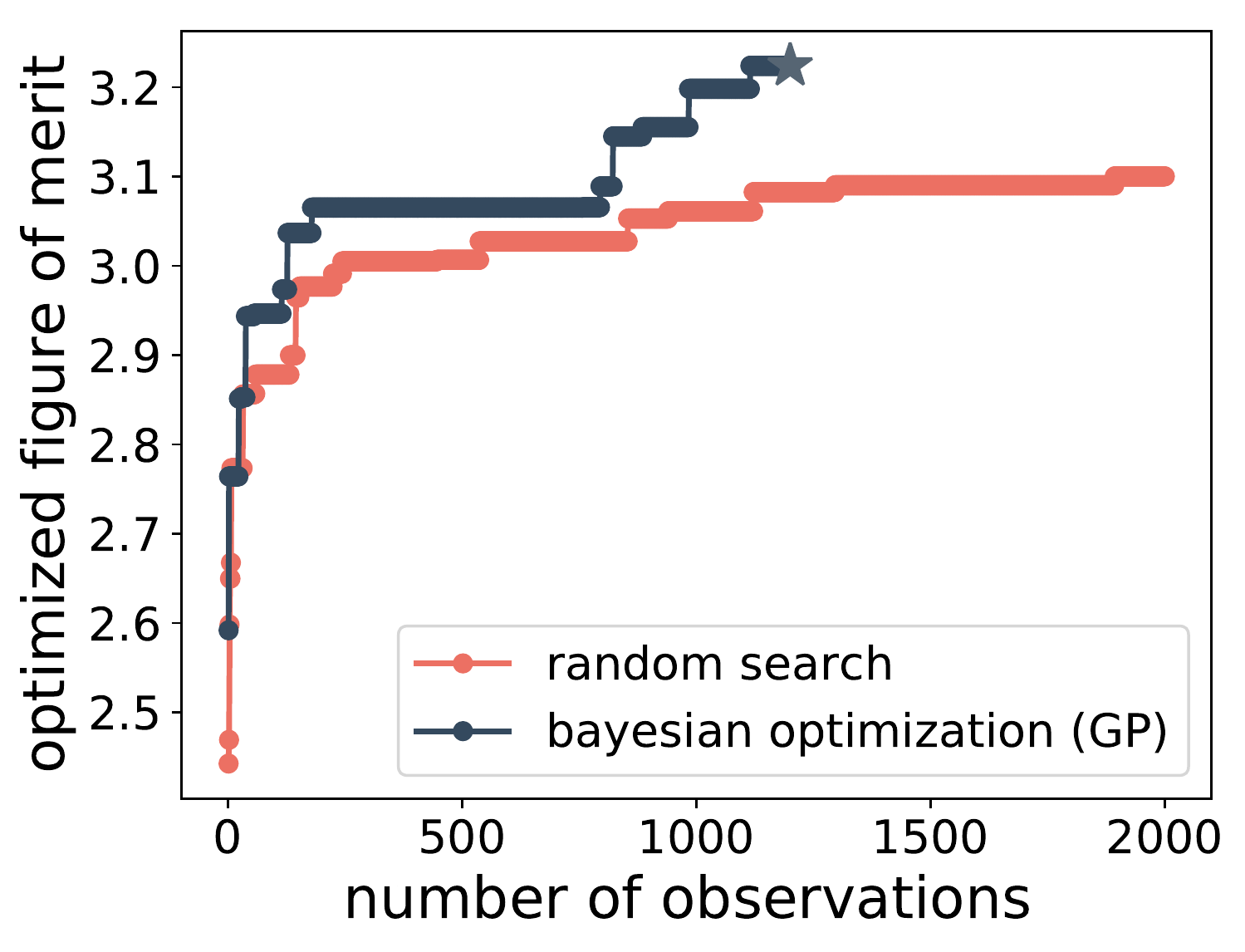}
\qquad
\includegraphics[width=.45\textwidth,origin=c,angle=0]{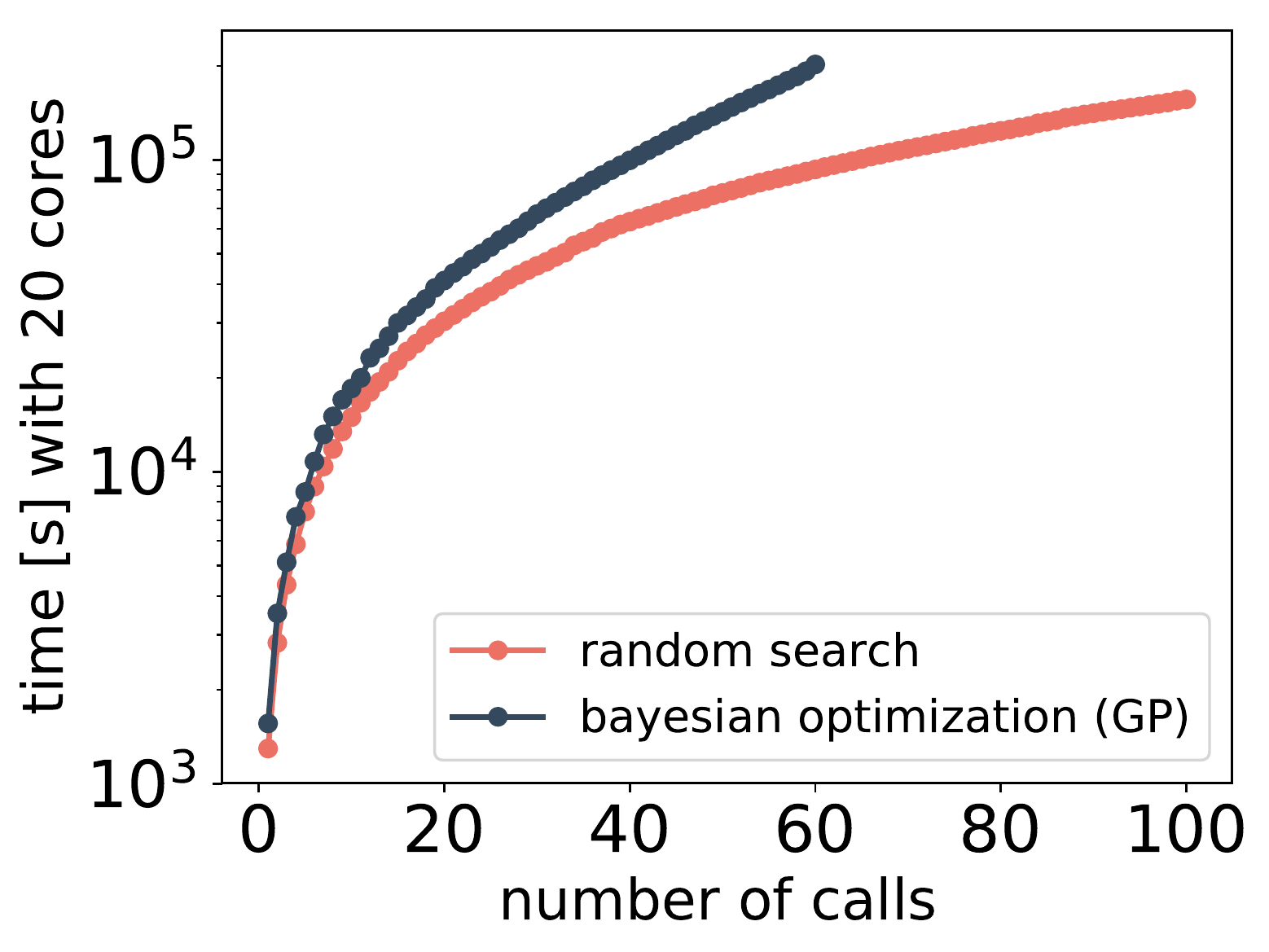}
\caption{\label{fig:comparison_fom_time} (left) Comparison of the Bayesian optimization with a random search. The random search after 2$\cdot$(10$^{3}$) observations ($T \times M$, see Table~\ref{tab:hyper_settings}) is far from the optimum, whereas the BO converged to the minimum within the tolerances. The star marker shows the point of minimum found by the BO. (right) The computing time as a function of the call $T$.
To make a consistent comparison between BO and RS, we use the same number of CPU cores for both approaches.
In contrast to the random search, the BO (with GP) scales approximately cubically with the number of observations.} 

\end{figure}

The search for the optimal FoM with the BO is compared to a standard random search (RS) in Fig.~\ref{fig:comparison_fom_time} (left).
A simple optimization technique like random search can be fast but as pointed out by Ref.~\cite{bergstra2012random} sequential model-based optimization methods---particularly Bayesian optimization methods---are more promising because they offer principled approaches to weighting the importance of each dimension.
Figure~\ref{fig:comparison_fom_time} (right) shows a comparison of the CPU time required for the two procedures. 

In both cases, to make a comparison using the same computing resources, the exploration of the parameter space at each call has been distributed  among $M$(=20) physical cores each evaluating a different point of the design space.\footnote{It should be obvious that the duration of each call for a random search is independent of the number of simulations running in parallel on different cores.}  
As the well known main drawback of the BO is that it scales cubically with the number of explorations, which is mainly due to the regression of the objective function through GPs (in Sec.~\ref{sec:future}, we discuss an improved strategy). Despite this, it typically converges to the optimum in a smaller number of iterations (and as the number of observations increase, it remains in the basin of attraction). %
A list of hyperparameters used in this framework can be found in Table~\ref{tab:hyper_settings}.

\begin{figure}[!t]
\centering 
\includegraphics[width=.65\textwidth,origin=c,angle=0]{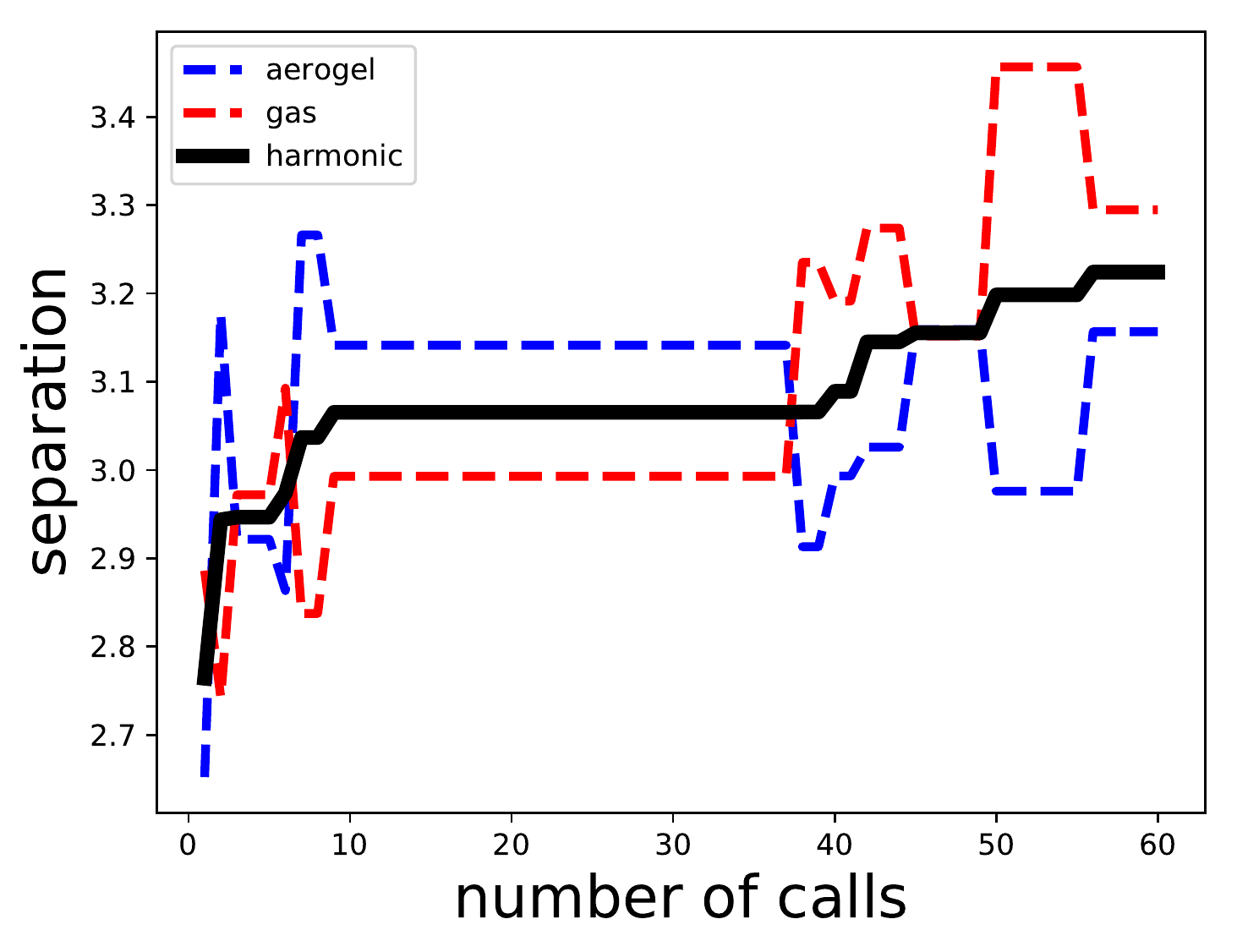} 
\caption{\label{fig:trade-off_drich} 
The best value of the figure of merit (black) as a function of the number of calls. The corresponding number of $\sigma$ for the low (blue, aerogel) and high (red, gas) momenta regions is also shown.
After a few calls where all curves increase, the optimization of the figure of merit is characterized by a trade-off between aerogel and gas, that is better performance in the gas implies a lower distinguishing power in the aerogel.
}
\end{figure}

Table~\ref{tab:final_separation} summarizes the results of the optimization procedure based on the figure of merit in Eq.~\eqref{eq:fom}. 
Figure~\ref{fig:trade-off_drich} shows the trade-off between the two regions in terms of PID performance found during the optimization process, where initially both aerogel and gas increase in separation power, and eventually after a certain number of calls a gain in the performance of the gas corresponds to a loss in the performance of the aerogel part.
Another interesting feature is suggested by the results of Fig.~\ref{fig:posterior}: 
we can increase the refractive index of the aerogel only if we increase its thickness and maintain a sufficiently high enough yield of Cherenkov photons.\footnote{The objective function $N\sigma$ (Eq. \ref{eq_nsigma}) at a given momentum is a trade-off between larger angular separation of two particle species (corresponding to lower refractive index) and larger number of photons (larger index of refraction and longer radiator). The number of produced photons will reach a plateau at some thickness value, due to the optical absorption length of the radiator material.} 
It is also relevant the fact that some of the geometrically constrained parameters, as well as the aerogel related ones, reached their range boundaries; the dRICH development shall take into account these results.


\begin{table}[t]
\caption{\label{tab:hyper_settings}
List of hyperparameters used in the optimization. {\em N.b.}, the maximum number of design points (observations) is T$\times$M, while the events simulated on each call are M$\times$N = 4000 for each particle species.
} 
\centering
\scalebox{0.8}{
\begin{tabular}{ | c | c | c |}
\hline
symbol & description & value \\
\hline
T & maximum number of calls & 100\\
M & design points generated in parallel & 20\\
N & pions (and kaons) per sample  & 200 \\
kappa & controls variance in predicted values & 1.96\\
xi & controls improvement over previous best values & 0.01\\
noise & expected noise (relative) & 2.5\%\\
\hline
\end{tabular}
}
\end{table}
\begin{table}[b!t]
\caption{\label{tab:final_separation}
Summary of the results obtained with the BO using GP. The number of $\sigma$ relative to the low momentum (aerogel) and high momentum (gas) regions correspond to the optimal value found for the figure of merit described in Eq.~\eqref{eq:fom}. 
}
\centering
\scalebox{0.8}{
\begin{tabular}{ | c | c | c | c |}
\hline
 & FoM (h) & $(N\sigma)$ @ 14 GeV/c  & $(N\sigma)$ @ 60 GeV/c \\
\hline
BO & 3.23 & 3.16 & 3.30 \\
legacy & 2.9 & 3.0 & 2.8 \\
\hline
\end{tabular}
}
\end{table}

\begin{figure}[!]
\centering
\includegraphics[scale=0.22, angle = 0]{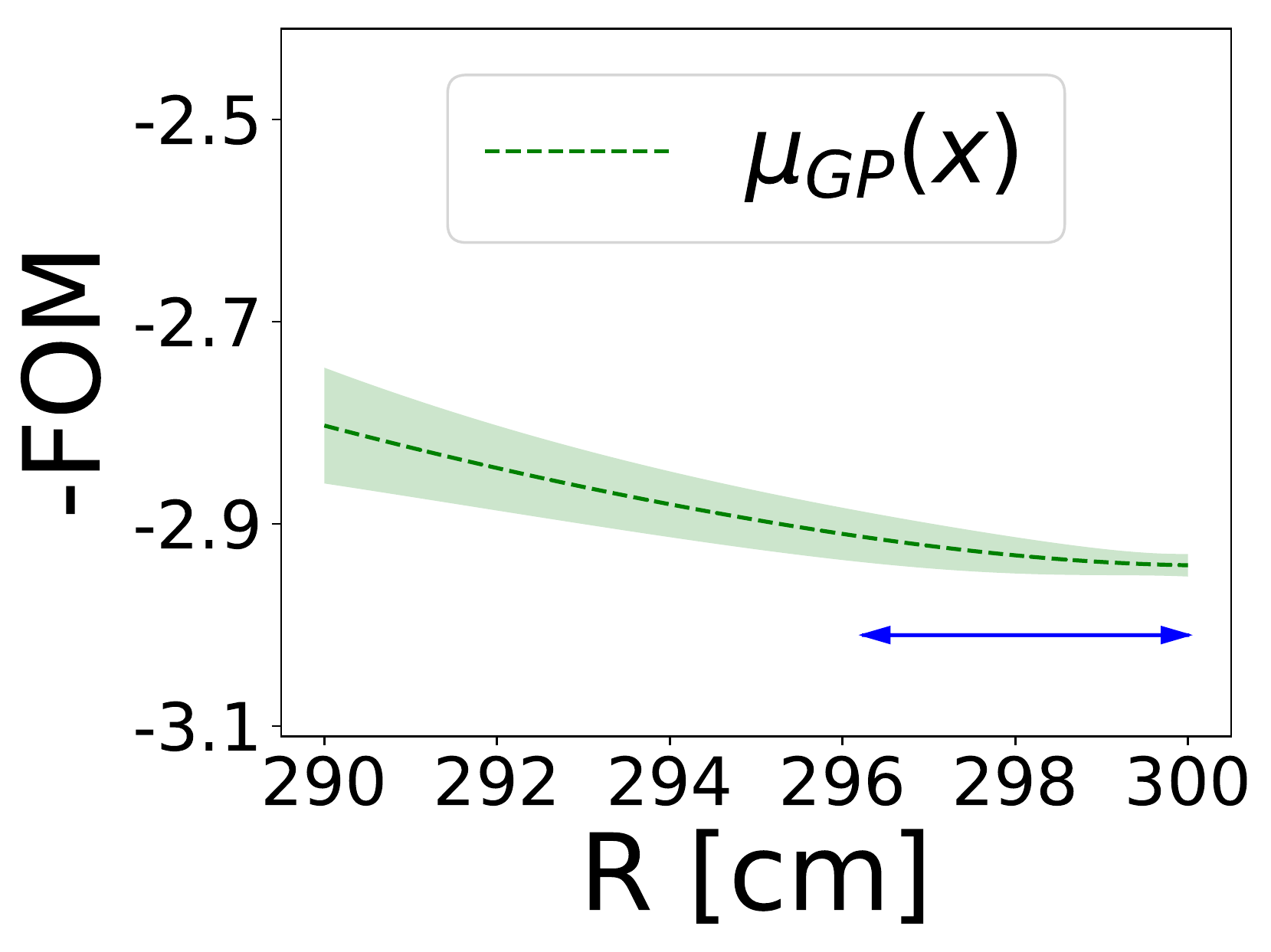}
\includegraphics[scale=0.22, angle = 0]{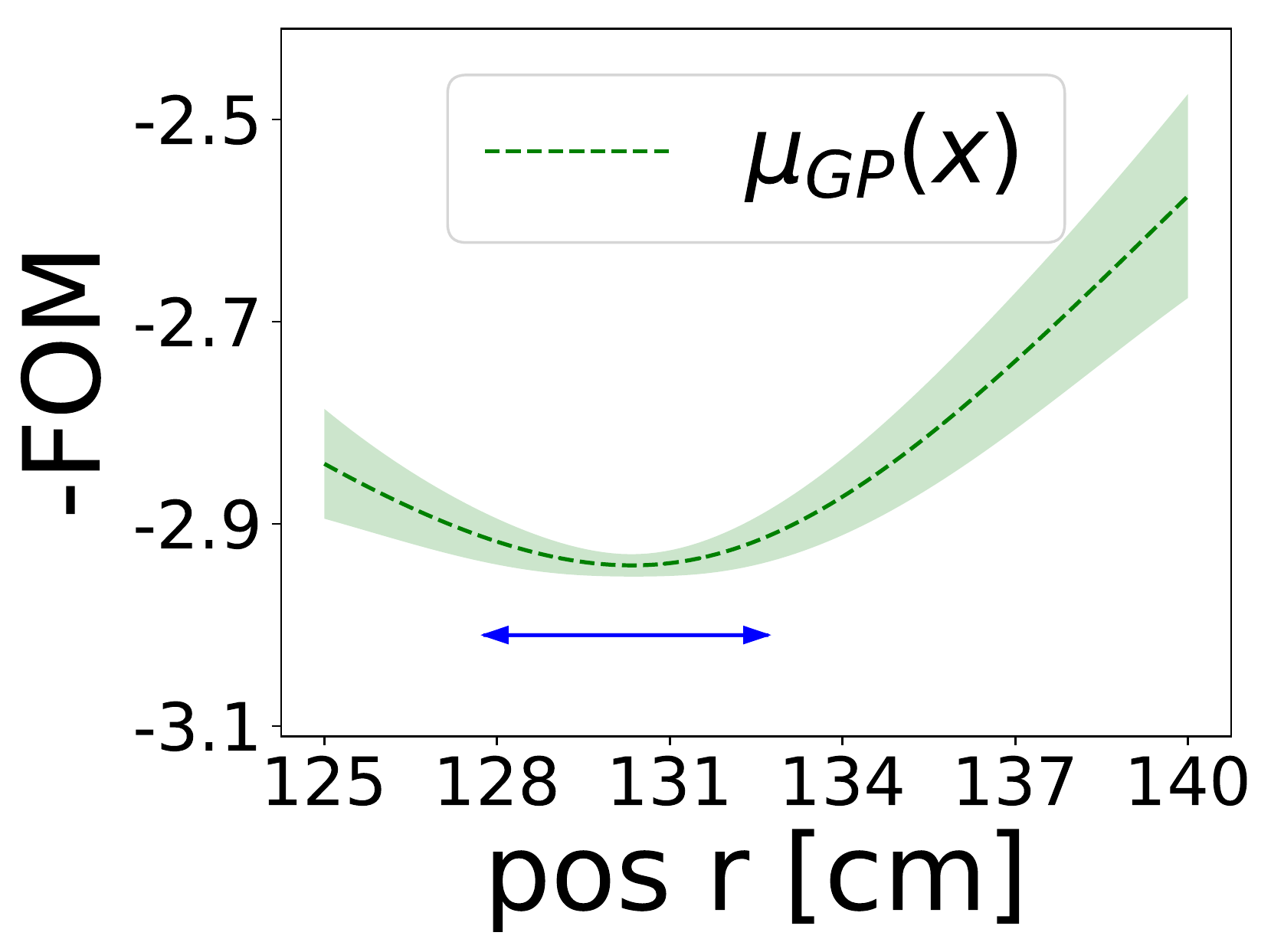}
\includegraphics[scale=0.22, angle = 0]{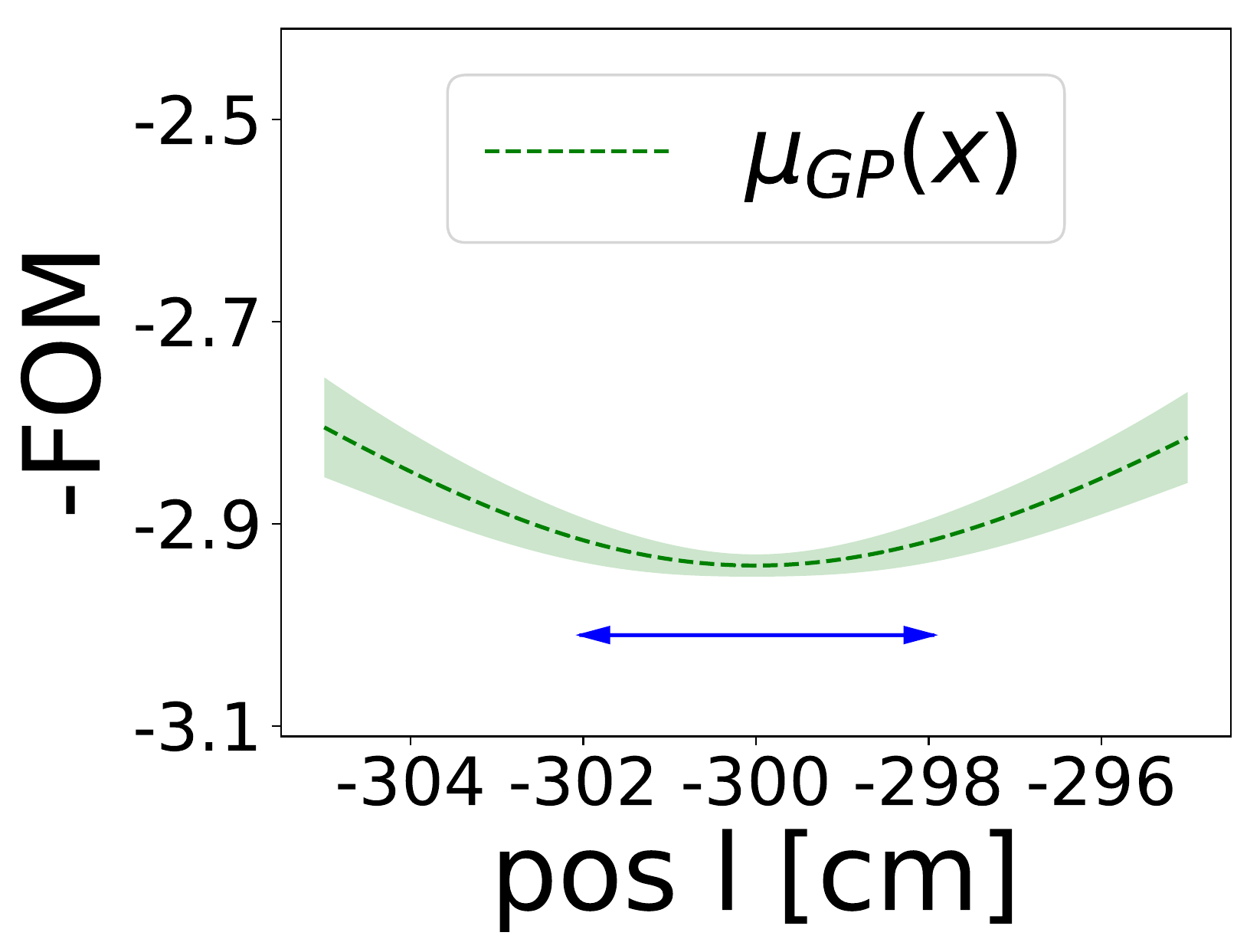}
\includegraphics[scale=0.22, angle = 0]{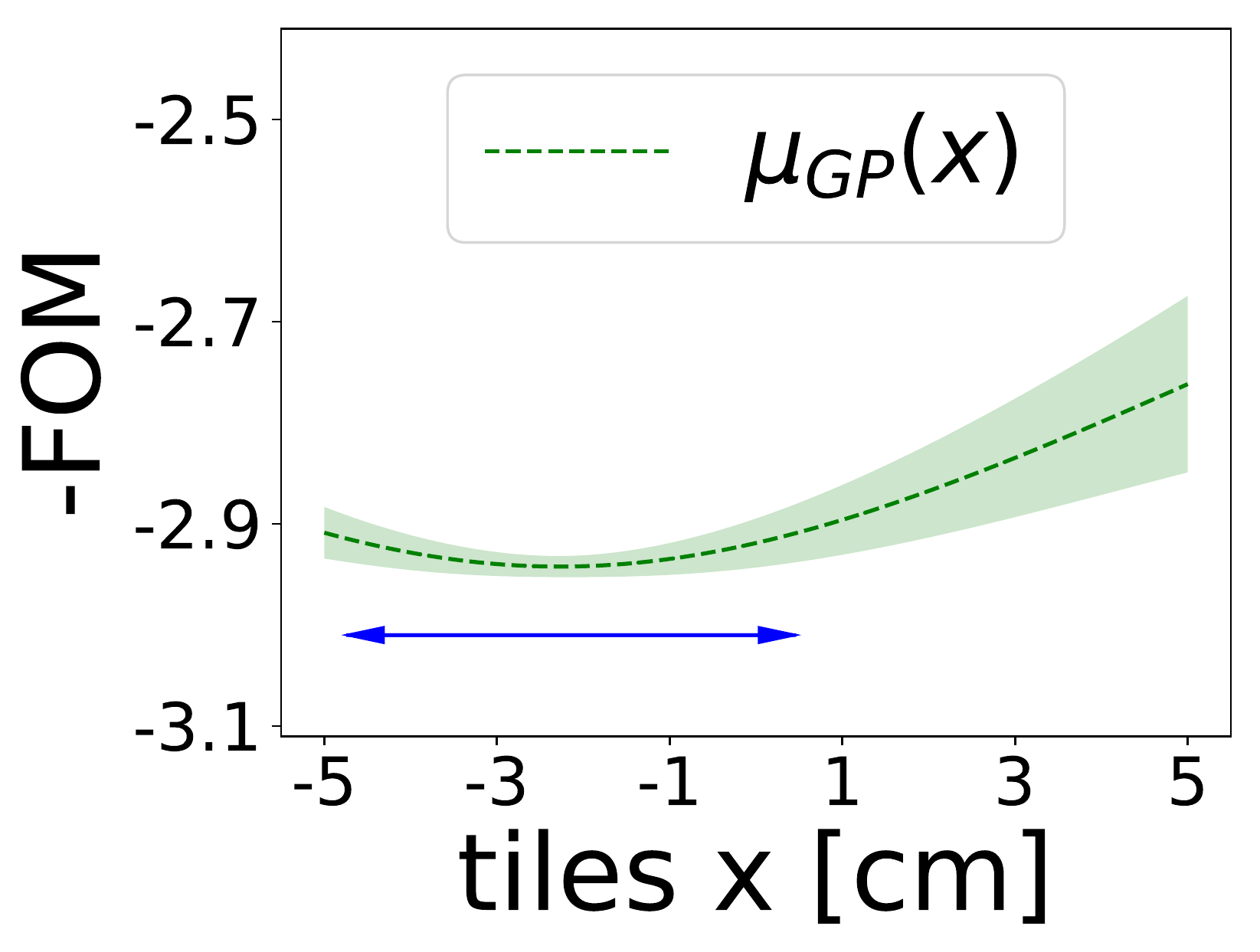}\\
\includegraphics[scale=0.22, angle = 0]{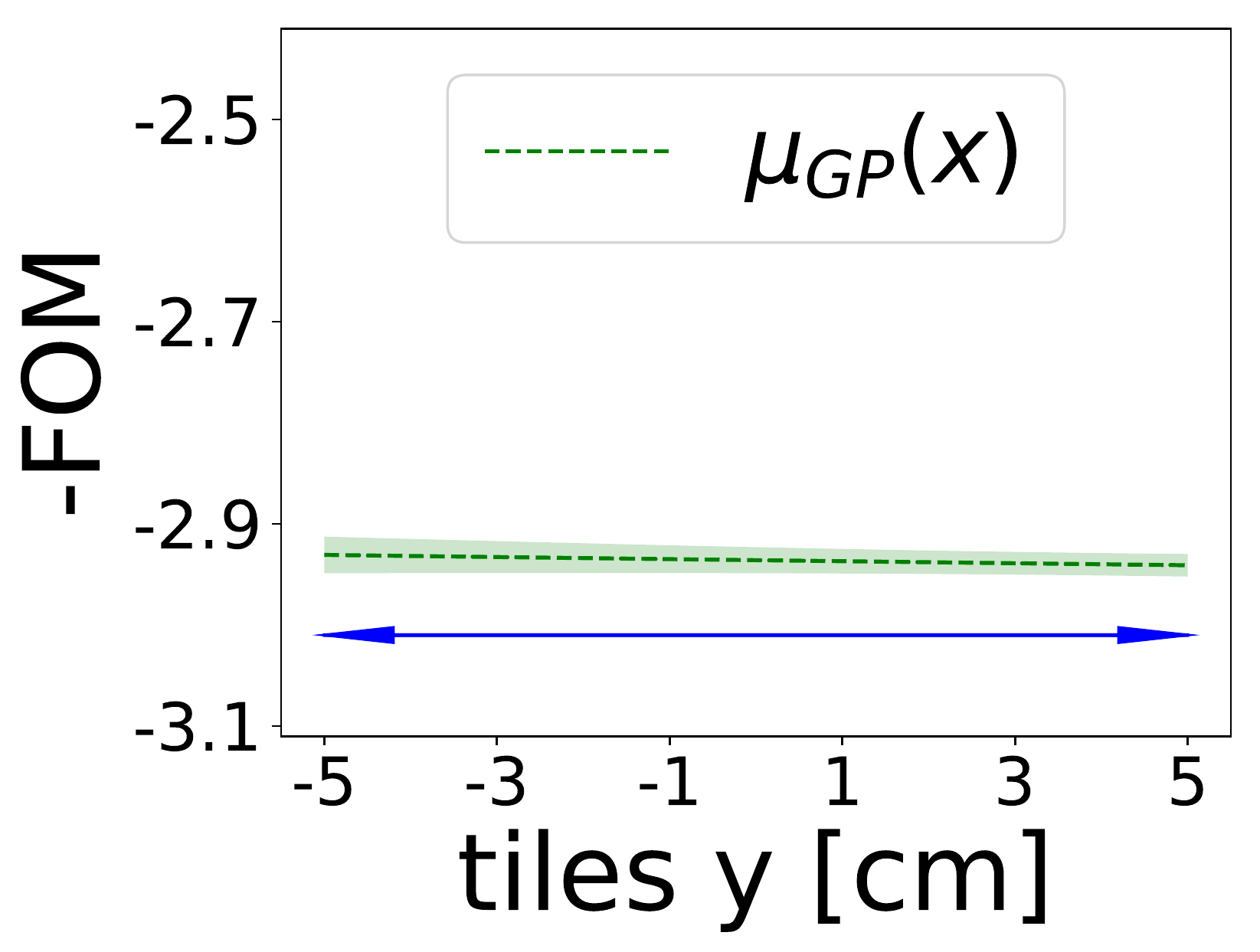}
\includegraphics[scale=0.22, angle = 0]{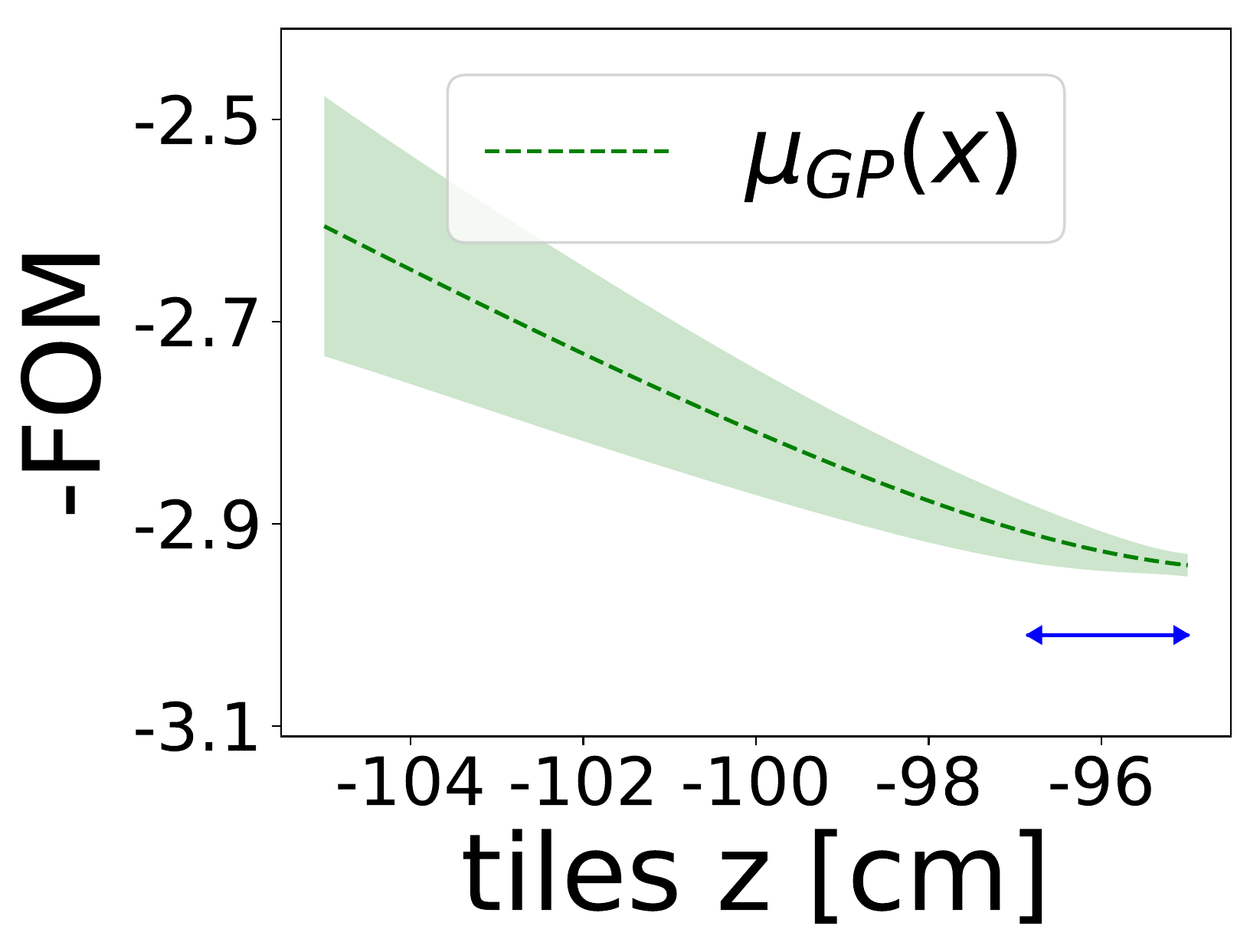}
\includegraphics[scale=0.22, angle = 0]{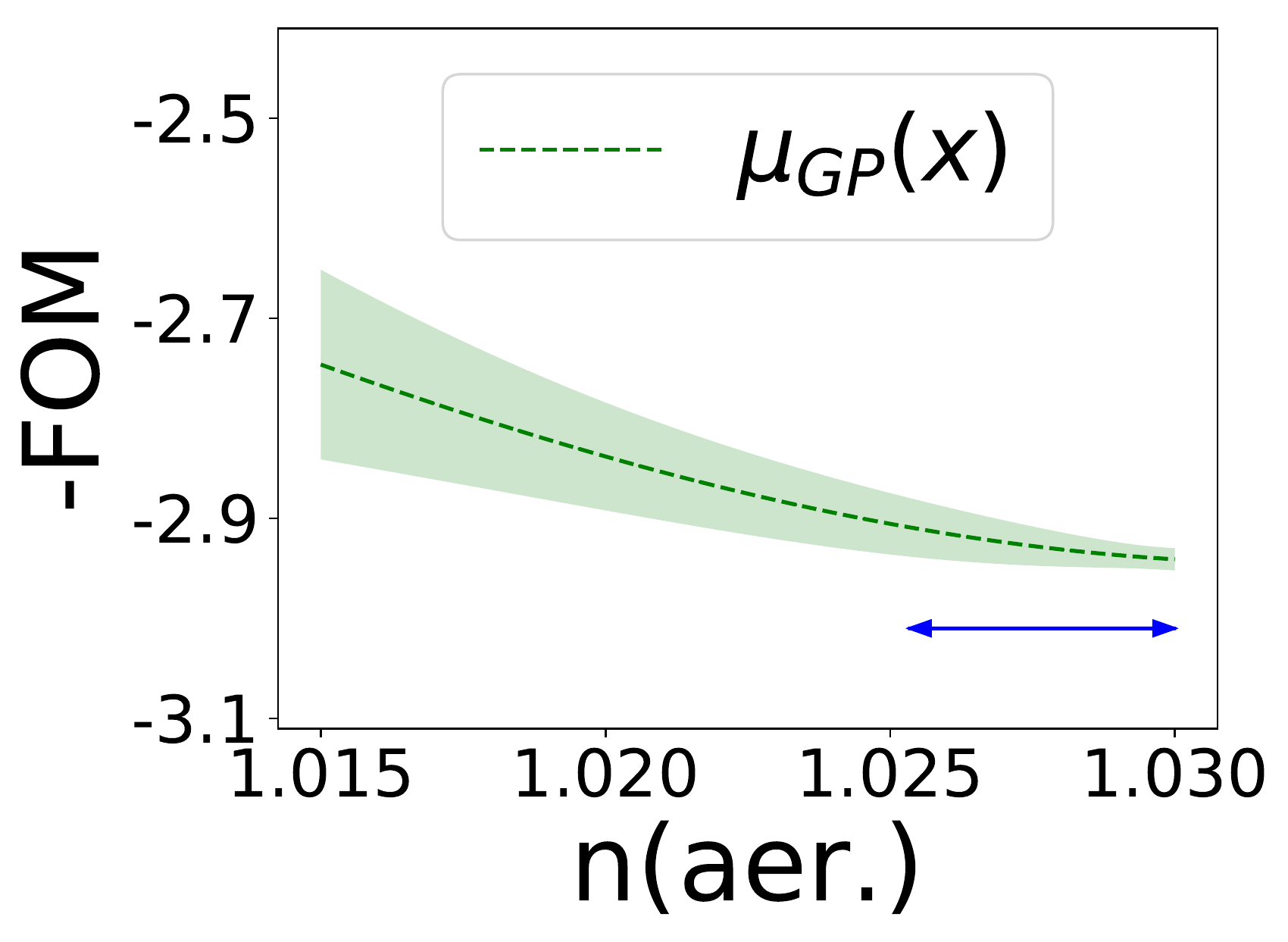}
\includegraphics[scale=0.22, angle = 0]{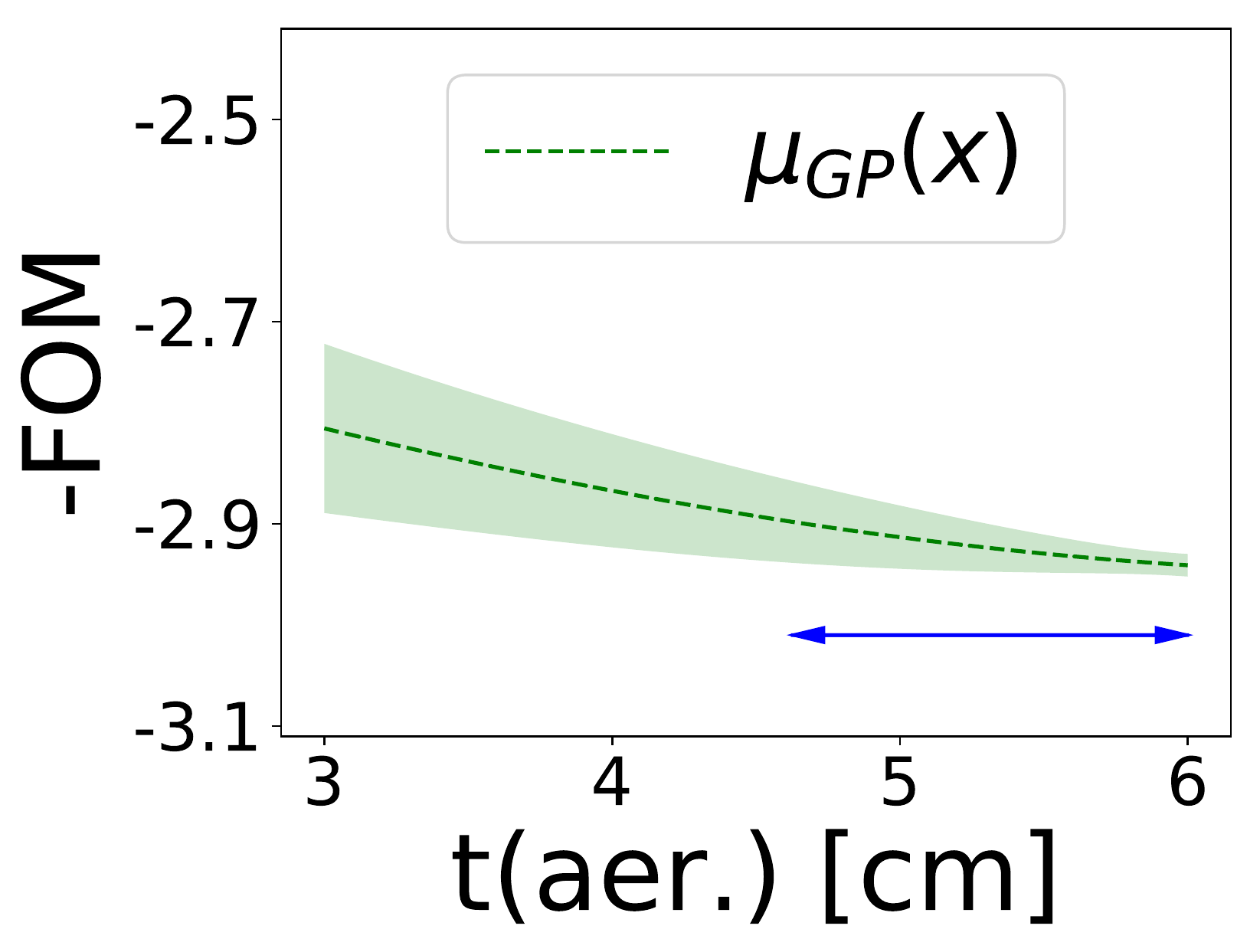}
\caption{The marginalized posterior distribution on each parameter obtained with BO. 
Notice that the FoM is defined with a negative sign. The two-headed blue arrow shows the tolerance region (as 95\% confidence level (C.L.) interval) on each parameter, where changing the parameter does not affect significantly the obtained performance. The behavior of the shift of these in the y direction (tiles y) is flat, indeed a small lateral shift of the tiles is expected to have no impact on the PID capability.
}
\label{fig:uncertainties}
\end{figure}

Since BO provides a model of how the FoM
depends on the parameters, it is possible to use the posterior to define a tolerance on the parameters.
Domain knowledge is required to decide how large of a change in the FoM is irrelevant, then the FoM space can be scanned to define how much each parameter can vary before seeing a relevant change. 
To this end, we use the proxy $Z = (\mu(x)-\mu(x_{opt})) / \sqrt{\sigma(x)^{2}+\sigma(x_{opt})^{2}}$, where $\mu(x)$ and $\sigma(x)$ are the expected value and standard deviation on the FoM as a function of $x$ provided by the model, and, in particular, $x_{opt}$ is the optimal value found by the BO for the considered parameter.  
The tolerance interval is defined as the region where $|Z| < 2$, this condition meaning a variation which is not statistically significant. 
The tolerance region of each parameter is drawn in Fig.~\ref{fig:uncertainties} as a two-headed blue arrow on top of the corresponding marginalized posterior distribution obtained with BO.
The values reported in Table~\ref{tab:tuned_parameters} are obtained following the above strategy.  

\begin{table}[!]
\caption{\label{tab:tuned_parameters}
The parameters of Table~\ref{tab:parameters} tuned by the Bayesian optimization, and the tolerance regions estimated according to the explanation of Sec.~\ref{sec:methodology}. } 
\centering
\scalebox{0.8}{
\begin{tabular}{ | c | c | c |}
\hline
parameter & results & tolerance region\\
\hline
 R & 300.0 & (296.2,300.0) [cm]\\ 
 pos r & 130.5 & (127.8,132.7) [cm]\\ 
 pos l & -299.8 & (-302.0,-298.0) [cm]\\   
 tiles x & -2.8 & (-4.7,0.4) [cm]\\
 tiles y &  5.0 & (-5.0,5.0) [cm]\\ 
 tiles z & -95.00 & (-96.85,-95.00) [cm]\\   
 n$_{aerogel}$ & 1.030 & (1.025, 1.030)\\
 t$_{aerogel}$ & 6.0 & (4.6, 6.0) [cm]\\
\hline
\end{tabular}
}
\end{table}

\begin{figure}[!]
\centering 
\includegraphics[width=.65\textwidth,origin=c,angle=0]{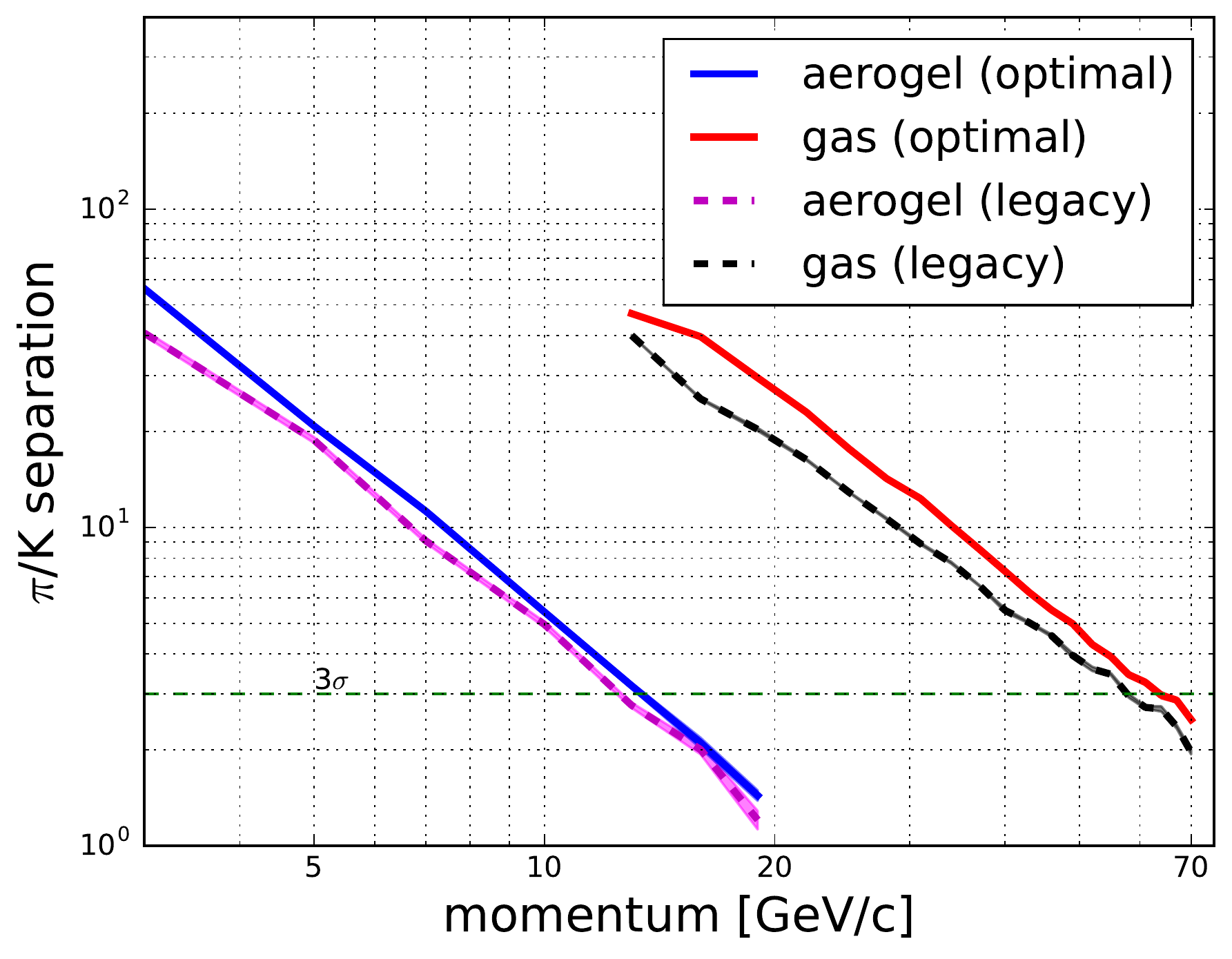}
\caption{\label{fig:improvement_baseline} 
$\pi/K$ separation as number of $\sigma$, as a function of the charged particle momentum. 
The plot shows the improvement in the separation power with the approach discussed in this paper  compared to the legacy baseline design \cite{del2017design}. The curves are drawn with 68\% C.L. bands which are barely visible in the log plot, but this lets better appreciate the significant difference between optimized and baseline curves.  
}
\end{figure}

The results discussed in the previous sections for the dRICH are very promising: Fig. \ref{fig:improvement_baseline} summarizes the overall improvement in PID performance over a wide range of momentum obtained with the Bayesian optimization compared to the baseline curve based on \cite{del2017design}. 
\\At least 3$\sigma$ $\pi/K$ separation is achieved in the aerogel region for $P \leq 13.5$ GeV/c (compared to the baseline 12.5 GeV/c), whereas in the gas region the same separation is obtained for $P \leq 64$ GeV/c (compared to 58 GeV/c). 

Both the baseline and the optimized curves have been calculated using charged tracks in the angular region (5,15) deg. 
As expected, once the optical properties of the aerogel are fixed (i.e. the refractive index dispersion curve, strictly connected with the single photon chromatic error), there is less room for a purely geometrical optimization of the system compared to the gas where the emission error is the dominant one and it is instead purely geometric.

\section{Conclusions and Future Directions}\label{sec:future}

We present for the first time an implementation of BO to improve the design of future EIC detectors, and we provide a detailed procedure to do this in a highly parallelized and automated way.
We show in this case study that the PID capabilities of the current modeled dRICH detector can be substantially improved using our approach.
In addition, the BO offers interesting hints on the relevance and correlation of the different parameters and it is possible to estimate the expected tolerances, within which any variation of the parameters does not alter the detector performance.
We expect that the proposed BO method can provide  improvements and hints for the future R\&D of the dRICH, when results from prototype tests will validate and consolidate the current simulated model.
More generally, real-world costs of the components could be included in the optimization process, either by extending the FoM or by exploring a Pareto optimization with multiple objective functions~\cite{ngatchou2005pareto}.

Currently, there are many ongoing efforts to simulate and analyse EIC detector designs, and the approach developed here---being completely general---can be employed for any such study.  
There are a variety of ways in which this study could be improved, which will be investigated in the near future. 
For example, in Appendix~\ref{sec:early_stop}, we introduce criteria for determining when to stop the optimization procedure, to avoid wasting resources if a suitable optimum has already been found. 
This work does not take into account the possible interplay between the dRICH and the other detectors. There is work ongoing on designing the TOF detector at very good timing resolution, and this will have an impact on the PID performance at low momentum. Therefore, a multi-detector design optimization is a possible future direction of this work.  
Another important aspect involves the choice of optimizing all the parameters together versus in blocks. 
For example, in the {\sc Pythia} BO tune~\citep{ilten2017event}, better precision was obtained by optimizing the parameters in 3 blocks rather than all of them together. This worked because many observables were unaffected by several {\sc Pythia} parameters. 
Once multi-detector optimizations are considered, it may be prudent to adopt a similar approach here. 
In addition, one can certainly test different figures of merit; however, the figure of merit defined in Eq.~\eqref{eq:fom} is well suited to optimizing the dRICH design, see Appendix \ref{app:dispersion}.  
Novel Python frameworks for Bayesian optimizations like GPflowOpt \cite{knudde2017gpflowopt} could improve the timing performance. 
This package is based on the popular GPflow library for Gaussian processes, leveraging the benefits of TensorFlow~\citep{abadi2016tensorflow} including parallelization.
Recently, optimization packages have been developed to rely on accelerated CUDA/GPU implementations. 
For example, Deep Networks for Global Optimization (DNGO)~\citep{snoek2015scalable}, where neural networks (NNs) are used as an alternative to GPs resulting in an improved scalability from cubic to linear as a function of the number of observations needed to explore the objective-function.   
The choice of hyperparameters of the NN is not defined {\em a priori} and in principle this could also be optimized with BO techniques to maximize the validation accuracy and inference time.

In conclusion, AI-based tools have been introduced and tested for the optimization of the dRICH configuration; the preliminary results clearly show a substantial improvement in performance and may provide useful hints on the relevance of different features of the detector. 
These same tools can be extended and applied to other detectors and possibly to the entire experiment, making the EIC R\&D one of the first programs to systematically exploit AI in the detector-design phase. 

\acknowledgments
We thank R. Yoshida for encouraging this work. This material is based upon work supported by the U.S.
Department of Energy, Office of Science, Office of Nuclear Physics under contracts DE-AC05-06OR23177, DE-FG02-94ER40818.
For this work, CF was supported by the EIC fellowship award at Jefferson Lab. 


\clearpage
\appendix
\section{Angular dispersion in RICH detectors} \label{app:dispersion}

The error on the Cherenkov angle is given by several contributions, $\sigma_{\theta_c}^{x_i}\equiv \frac{\partial \theta_c}{\partial x_i}\sigma_{x_i}$; the error on the measured Cherenkov angle of a single photon is given by the quadratic sum of all the relevant contributions:
\begin{equation}
\label{quad_sum}
    \sigma_{\theta_c}=\sqrt{\sum_{x_i} \left (\frac{\partial \theta_c}{\partial x_i}\sigma_{x_i}  \right )^2}
\end{equation}
The single terms in the sum that we consider are the typical relevant contributions taken into account for similar dual-radiator RICHes in spherical reflecting mirror configuration (see \cite{nappi2005ring, ypsilantis1994theory} for a compendium). 
The reconstruction of the Cherenkov angle in Fig. \ref{fig:dual2} is based on the GEMC simulation and the indirect ray tracing algorithm used by the HERMES experiment
(see \cite{akopov2002hermes} for details on the algorithm).
The following contributions are taken into account in the simulation.

\paragraph{Chromatic error:} 
The most important aspect in the choice of a Cherenkov radiator for a RICH detector is the index of refraction of the material and its characteristic optical dispersion.
Namely, one of the terms in \ref{quad_sum} is due to the variation of the refractive index of the materials traveled by the photon of a given energy (or wavelenght $\lambda$). In the case of a dual-radiator (let us assume aerogel and gas) this error consists of two contributions for the first radiator: the uncertainty on the photon emission wavelength depending on the dispersion relation ($\sigma_{\theta_c}^{\lambda(dis)}$) and the uncertainty on the refraction between aerogel and gas surface ($\sigma_{\theta_c}^{\lambda(ref)}$). Therefore we have:
\be
\sigma_{\theta_c}^\lambda=\frac{d\theta_c}{d\lambda}\sigma_\lambda\sim \sqrt{(\sigma_{\theta_c}^{\lambda(dis)})^2+(\sigma_{\theta_c}^{\lambda(ref)})^2}
\ee 
where 
\be
\sigma_{\theta_c}^{\lambda (dis)}\equiv\frac{\partial \theta_c}{\partial n_a}\frac{dn_a}{d\lambda}\sigma_\lambda
\ee
with $n\equiv n(\lambda)$ the refractive index of the considered radiator.
//For the emission contribution to the chromatic error, we have
\be
\frac{d\theta_c}{d\lambda}=\frac{\partial \theta_c}{\partial n}\frac{dn}{d\lambda}=\frac{1}{n^2\beta\sin\theta_c}\frac{dn}{d\lambda}
\ee
and
\be
\sigma_{\theta_c}^{\lambda(dis)}=\frac{1}{n^2\beta\sin\theta_c}\frac{dn}{d\lambda}\frac{\Delta\lambda}{\sqrt{12}}
\ee

The above can be also found in \cite{nappi2005ring} using $E$ instead of $\lambda$.
Notice that the variation of the refractive index of the aerogel, namely $\frac{dn_a}{d\lambda}$, can be inferred by recent measurements on the latest generation aerogel (i.e. the one tested by the CLAS12 RICH collaboration \cite{pereira2016test}).
The uncertainty due to the refraction of the photons between the two materials, $\sigma_{\theta_c}^{\lambda(ref)}$, is in general one order of magnitude lower than the chromatic aberration; nevertheless it has to be considered and somehow corrected in any kind of geometric reconstruction algorithm.  
For the second radiator (the gas in this case) only the first contribution is present. 
The chromatic error curves have been obtained using the following optical parameters:
\begin{itemize}
    \item aerogel $n(\lambda)$ and transmittance: the detailed study of the CLAS12 RICH collaboration has been used to infer the optical properties of recently manufactured aerogel \cite{pereira2016test}. Also the scattering length, and the related Rayleigh scattering have been introduced in the GEMC simulation. The photons with wavelength below 300 nm have been filtered simulating a thin acrylic slab placed between aerogel and the gas tank, for both shielding and avoiding aerogel chemical degradation;
    
    \item $C_2F_6$ properties from \cite{abjean1995refractive};
    
    \item the mirror reflectance has been assumed the same of CLAS12 HTCC RICH mirrors (this is the GEMC mirror reference \cite{gemc}, the reflectance varies in the range of wavelength between 190 and 650 nm, consistent with the photon sensor quantum efficiency); 
    
    \item the quantum efficiency curve of the multi-anode PMT Hamamatsu-H12700-03 \cite{h12700}.
\end{itemize}

\paragraph{Emission error:} The focal plane shape has been chosen to be spherically-tessellated (about 4500 cm$^2$ per sector, $5\times 5$ cm$^2$ each tile). 
Due to the spherical aberrations the real focal plane for a spherical mirror is no longer a naive
sphere. 
Therefore, the real shape of the focal surface will depend, point by point, on the angle (with respect to the optical axis) of incidence of the photon track on the mirror.
Namely, this angle is known only statistically due to the fact that the emission point of each photon in the radiators is unknown.

\paragraph{Pixel-size error:} This corresponds to the uncertainty due to the granularity of the pixel detector, namely 3 mm pixel size in dRICH case. 

\paragraph{Magnetic field error:}
The dRICH is immersed in a non negligible magnetic field and the charged hadron tracks are bending as they pass through the Cherenkov radiators (and expecially the much longer gas).
This introduces
an additional source of error in the Cherenkov angle. The effect is proportional to the path
length within the Cherenkov radiators, and therefore it becomes particularly important for
the gas radiator at large polar angle. 

\paragraph{Track error:} An angular track smearing of 0.5 mrad (1 mm over 2 m) has been assumed with very safe margin of technical feasibility.

\section{Noise Studies} \label{sec:noise_studies}

We investigated and characterized the statistical noise present on the objective function during the optimization process. 
Notice that Bayesian optimization takes into account this noise as an external parameter, e.g. skopt \cite{skopt} allows to deal with this noise in different ways, for example by adding it (assumed to be ideal gaussian) to the Mat\'ern kernel \cite{williams2006gaussian}.
From simulations we have been able to determine the relative uncertainty on the figure of merit, which as expected scales $\propto 1/\sqrt{N_{track}}$ (e.g. $\sim$ 2\% if $N_{track}=400$).

\begin{figure}[!]
\centering %
\includegraphics[width=.90\textwidth,origin=c,angle=0]{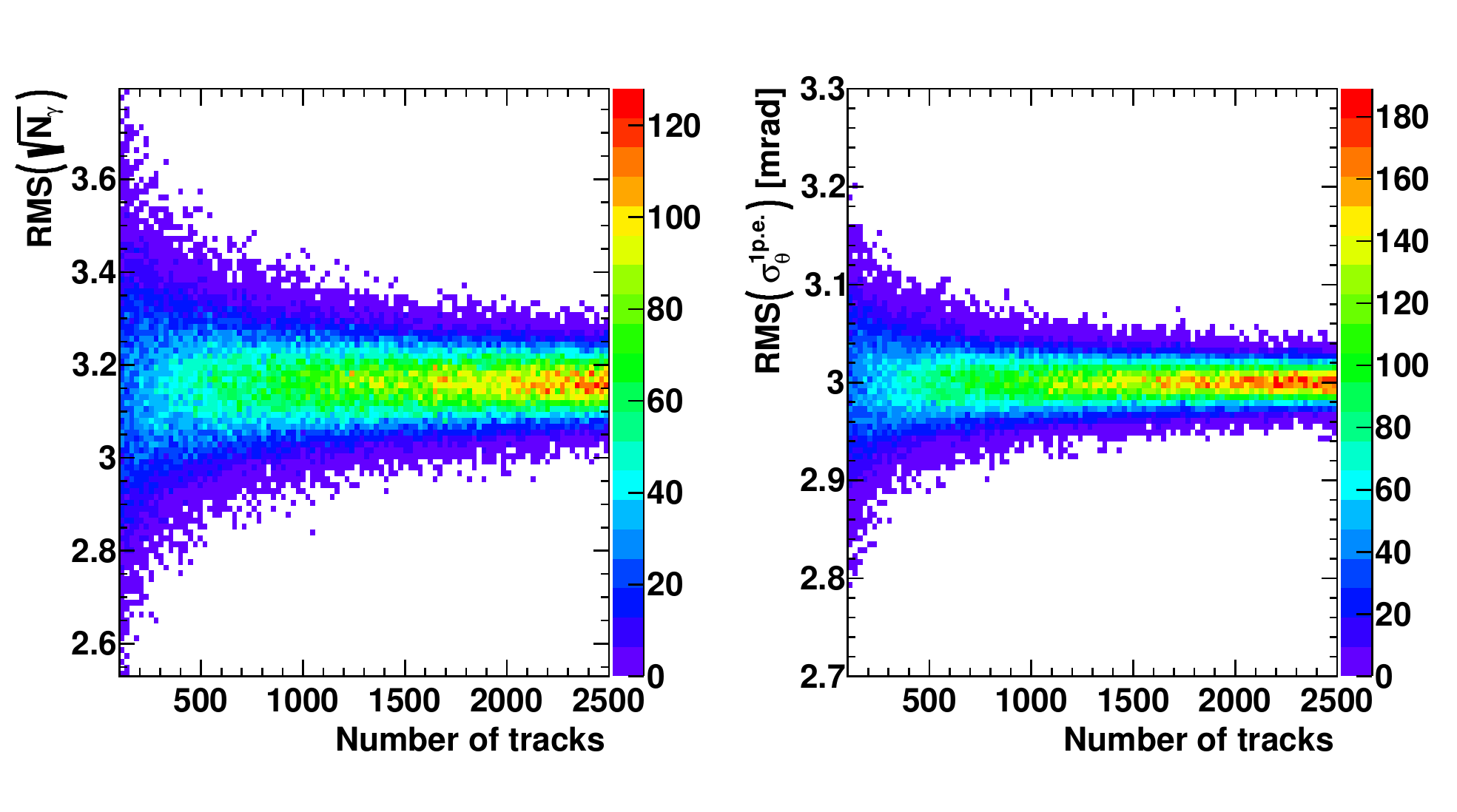}
\caption{\label{fig:app_uncertainty1} 
Uncertainties vs the number of tracks for aerogel assuming 10 photoelectrons and 3 mrad single photon angular resolution. Left: uncertainty on $\sqrt{N_\gamma}$; right: uncertainty on $\sigma_\theta ^{1p.e.}$ (in mrad unit).
}
\end{figure}

\begin{figure}[!]
\centering %
\includegraphics[width=.90\textwidth,origin=c,angle=0]{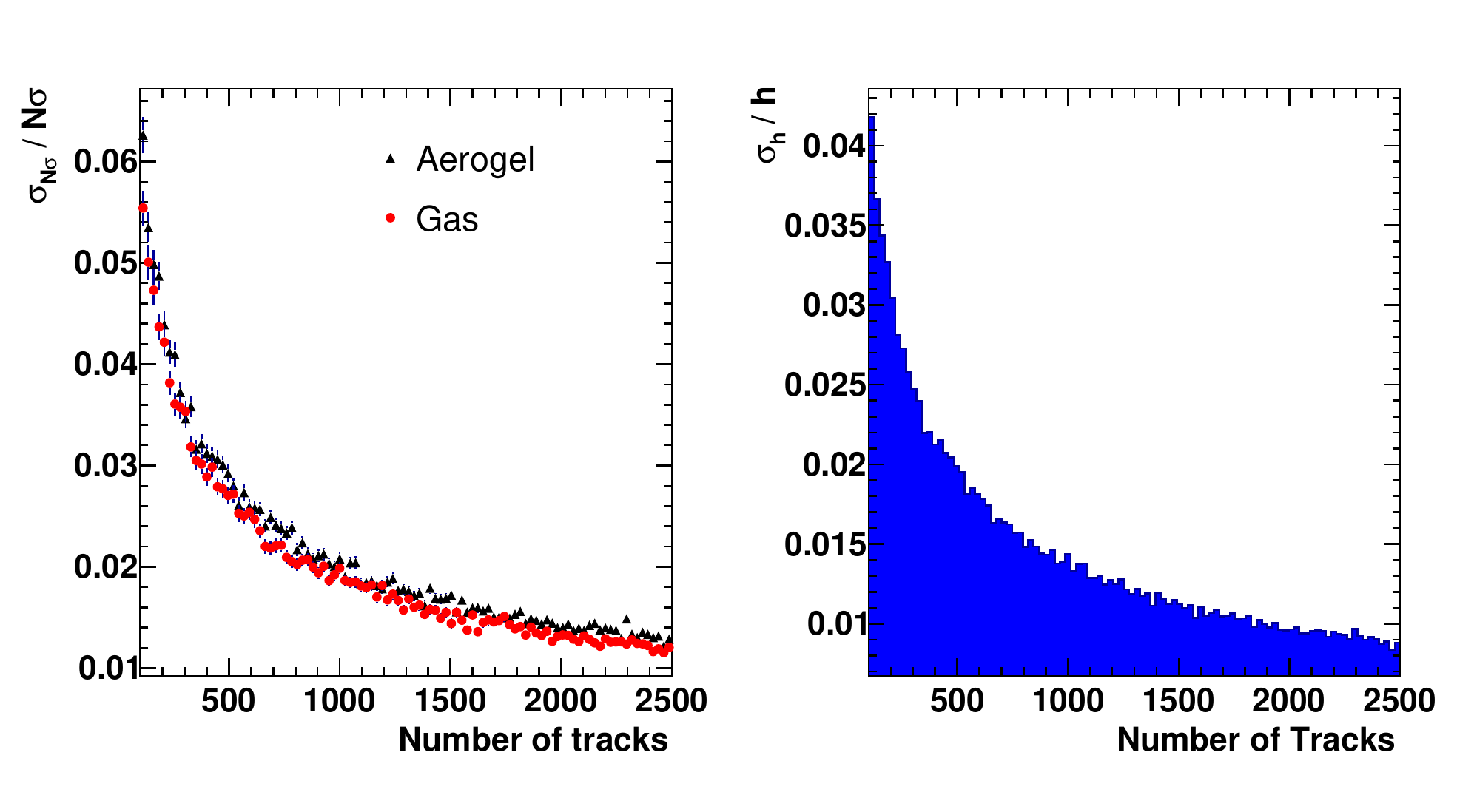}
\caption{\label{fig:app_uncertainty2} 
Uncertainties vs the number of tracks: (left) Relative uncertainty on the resolution in the aerogel (red) and in the gas (black), cf. Eq. \eqref{eq_nsigma}. (right) Relative uncertainty on the harmonic mean defined in Eq. \eqref{eq:fom}.
}
\end{figure}

The relative fluctuations of the terms $\sqrt{N_{p.e.}}$ and $\sigma_{\theta}$ (Fig. \ref{fig:app_uncertainty1}) contribute roughly equally in Eq. \eqref{eq:fom} as one can expect from the statistical errors propagation; these fluctuations are largely independent on the absolute values of the above terms (as demonstrated by the left plot in Fig. \ref{fig:app_uncertainty2}) and therefore on the charged particle momenta (as long as they are above Cherenkov threshold). 

In the results shown in Fig. \ref{fig:app_uncertainty1} and \ref{fig:app_uncertainty2}, we neglected the uncertainty on the difference between the mean angles, which is smaller than the uncertainty on $\sqrt{N_{p.e.}}$ and $\sigma_{\theta}$.   

\vspace{5cm}

\section{Early stopping criteria}\label{sec:early_stop}


Each call is characterized by $m$ design points (generated in parallel), each being a vector in the parameters space defined in Tab. \ref{tab:parameters}, with $d$-dimensions. At each call $n$, we can calculate the average point (vector) from the $m$ vectors ${X_{k}}$ (where $k=1,\cdots,m$), as:

\begin{equation}\label{eq:mean_x}
    \bar{\vec{X}}^{(n)}=\frac{1}{m}\sum_{k=1}^{m}{\vec{X}_{k}}^{(n)}.
\end{equation}

For a large number of calls (after a burn-in period chosen to be equal to 10 calls), the design point expressed by Eq. \eqref{eq:mean_x} tends to a global optimum steered by the BO.

In the following we will use the notation $X_{k,i}^{(n)}$ to express the $i_{th}$ element of the $k_{th}$ point, where $i=1,\cdots,d$ and $k=1,\cdots,m$. 
At each call, we can also define the sample variance on each component of Eq. \eqref{eq:mean_x}, in the following way: 

\begin{equation}\label{eq:variance_x}
    s{^{2}_{i}}^{(n)}=\sum_{k=1}^{m}\frac{\left( X^{(n)}_{k,i}-\bar{X}^{(n)}_{i}\right) ^{2}}{m-1}
\end{equation}

Pre-processing of data consists in masking for each component the outliers distant by more than 3$\sigma$ from the average component. They correspond to very large explorations in the optimization process.   
Therefore, the effective number of points for each component and at each call becomes $m^{(n)}_{i} = m - N^{(n)}_{i}(outliers)$.
and we can redefine the masked average and variance expressed in Eq. \eqref{eq:mean_x}  \eqref{eq:variance_x} after removing these outliers.
To simplify the notation, we will dub the masked mean and variance on every component as $\bar{X}^{(n)}_{i}$ and $s{^{2}_{i}}^{(n)}$.\footnote{Standard deviations smaller than tolerances are replaced with the latter in Eqs. \eqref{eq:standardized_x},\eqref{eq:fisher_sx}.}
\begin{figure}[!ht]
\centering
\includegraphics[scale=0.45, angle = 0]{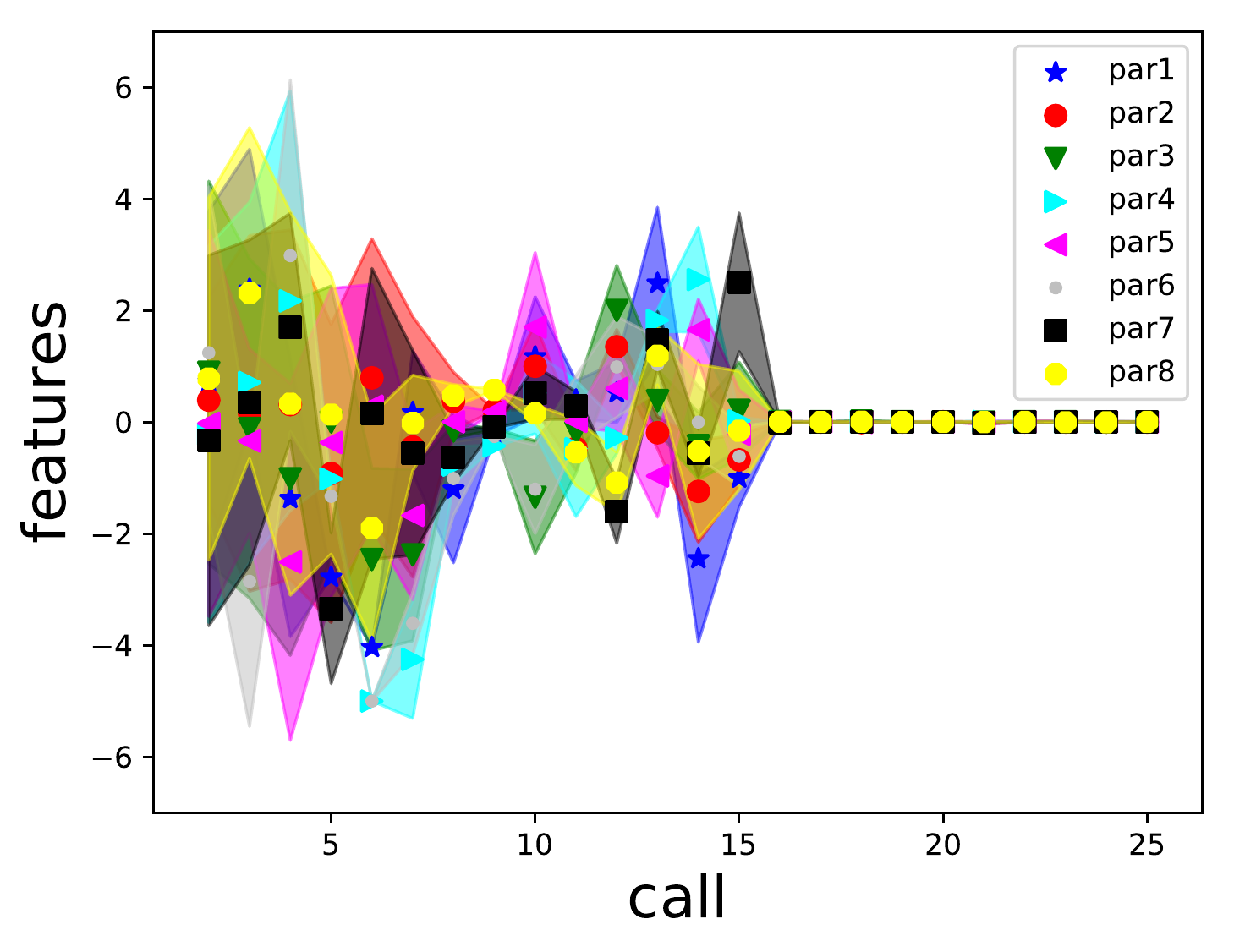}
\includegraphics[scale=0.45, angle = 0]{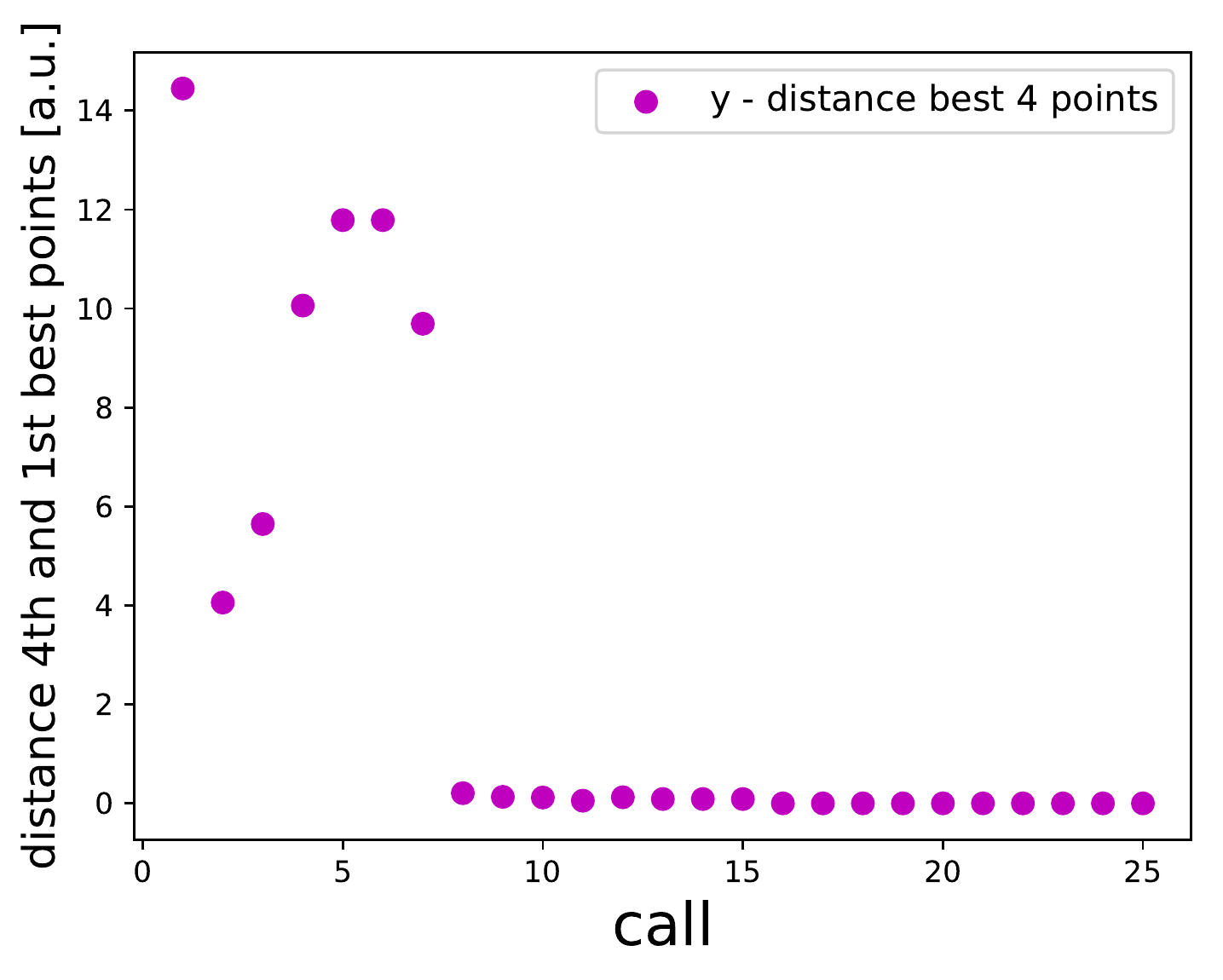}
\caption{The average (over M parallel cores) parameters chosen by the BO as a function of the number of calls. When the stopping criteria are satisfied the algorithm is interrupted. One can notice how the uncertainty on each average value is restricted when the above conditions are verified. For a simpler visualization, the results shown are obtained with a toy model $-$ 8 parameters tuned, sampling 20 points per call (the computation of each point is distributed on a different physical core as explained in the text).}
\label{fig:dual1x}
\end{figure}
This allows to define the standardized variable: 
\begin{equation}\label{eq:standardized_x}
    Z_{i}^{(n)} = \frac{\bar{X}^{(n)}_{i}-\bar{X}^{(n-1)}_{i}}{\sqrt{s{^{2}_{i}}^{(n)}+s{^{2}_{i}}^{(n-1)}}}
\end{equation}

Notice that Eq. \eqref{eq:standardized_x} is defined comparing the values in the previous ($n-1_{th}$) and current ($n_{th}$) calls.\footnote{Rigorously if $m<$40 typically one refers to small sampling theory and a t-distibution should be considered. The stopping criteria have been implemented in a flexible way that allows to use either a standardized Z or t-distribution depending on the number of physical cores available.}
Now at each call and for each component, we can test the null hypothesis $H_{0}$ that $\bar{X}^{(n)}_{i}$ and $\bar{X}^{(n-1)}_{i}$ belong to a population with same mean $\mu_{i}$, that is they are converging in the $i_{th}$ component to the same value. 
We do $d$ two-tailed p-value tests at 0.3\% significance level (corresponding approximately to $|Z^{(n)}_{i}|<$3) for each component ($i=1,\cdots,d$) in the parameter space. 

\be\label{eq:stop_x}
\chi^{(n)}(\bar{X}) = \Pi_{i=1}^{d} \chi^{(n)}_{Z}(X_{i})
\ee
The $\chi$-functions in Eq. \eqref{eq:stop_x} represent boolean operators, that is each $\chi^{(n)}_{Z}(X_{i})$ is 1 when the null $H_{0}$ is confirmed on the $i_{th}$ component. 
In addition, it is possible to build a Fisher statistic, 

\be\label{eq:fisher_sx}
F=s{^{2}_{i}}^{(n)}/s{^{2}_{i}}^{(n-1)},
\ee

which can be used to determine whether or not the variance $s{^{2}_{i}}^{(n)}$ is significantly larger than $s{^{2}_{i}}^{(n-1)}$ at a certain significance level, e.g., 1\%. 
This is done for all the components, similarly to what already discussed for \eqref{eq:standardized_x}, and enters as a multiplicative condition in \eqref{eq:stop_x}.
If all of them are 1 then the overall function $\chi^{(n)}(\bar{X})$ becomes 1 too (in that case meaning the \textit{x stopper is activated}).
The stopping criteria on $\vec{X}$ at the call $n_{th}$ corresponds to $d$ p-value tests all confirming the hypothesis $H_{0}$ as well.
In order to stop the BO, the other condition (\textit{y stopper is activated}) has to be satisfied along with Eq. \eqref{eq:stop_x}.

\be\label{eq:stop_x_all}
\chi^{(n)}(\bar{X}) = \Pi_{i=1}^{d} \chi^{(n)}_{Z}(X_{i}) \cap \chi^{(n)}_{F}(X_{i})
\ee

%
%
As for the value of the figure of merit $y$, we consider at each call the best $q$ optimal values (e.g. $q =5$) obtained up to that call, and calculate the difference between the worst $q_{th}$ (largest) and the best $1_{st}$ (smallest) as\footnote{In a minimization problem the lowest of the minima is considered the best value.} 

\begin{equation}\label{eq:delta_y}
    \Delta y^{(n)} = \underset{l=1,\cdots,q}{\max \{ {y}^{(n)}_{l} \} - \min \{ {y}^{(n)}_{l} \} }
\end{equation}

The statistic Eq. \eqref{eq:delta_y} vary typically rather smoothly if q$\leq$5, and we can apply a cut on the relative variation between two successive calls, requiring 

\be\label{eq:stop_y}
\textup{abs}(\Delta y^{(n)}-\Delta y^{(n-1)})/\overline{\Delta y} < 5\%,
\ee
where $\overline{\Delta y}$ is the arithmetic mean between $\Delta y^{(n)}$ and $\Delta y^{(n-1)}$.
The threshold values used to activate the above booleans have been chosen based on a toy model with the same number of parameters to tune. 
In conclusion, when the two conditions \eqref{eq:stop_x} and \eqref{eq:stop_y} are both activated or the maximum number of calls reached then the search is stopped.
It should be clear that the empirical stopping criteria expressed in Eq. \eqref{eq:stop_x} and \eqref{eq:stop_y} are not sufficient to exclude the presence of a local optimum. Nevertheless Bayesian optimization is one of the most suitable tools to find a global optimum in a bounded region and we made dedicated studies in Appendix \ref{sec:noise_studies} to check the consistency of the values of our cuts when the stoppers are triggered with the expectations (e.g.,we know that the relative statistical uncertainty on the harmonic mean for a number of tracks equal to 400 is about 2\%). 
We can also compare the results obtained using a regressor other than GP (e.g. GBRT, extra trees (ET), random forest (RF) regressors etc. \cite{skopt}), as explained in the text.
A simple random search is fast enough to run and can provide useful hints if the candidate point is far from the optimum.

Notice that \eqref{eq:standardized_x} could produce potential issues if the RMS in the denominator is large compared to numerator.  
The following combined requirements prevent from these issues: (a) minimum number of burn-in calls, (b) all the booleans are true in Eqs. \eqref{eq:stop_x_all} and \eqref{eq:stop_y}, (c) additional request that the above conditions (a) and (c) are true for a call and check that this holds in the successive call before activate the early stopping. 

Another parameter to define is the maximum number of calls that stops the search in case the other stopping criteria did not trigger. 
We consider the heuristic formula $\sim 25 \cdot n(pars)$ determined by \cite{ilten2017event} as a possible lower bound.\footnote{An hand-waving estimate for the number of calls needed when running simulations in parallel is obtained by scaling that number by $n(cores)$.}
Clearly the total number of calls can be set to any larger value provided enough computing/time resources. 


\bibliography{mybibfile}

\end{document}